%% file: bvp_longPaper.tex
\documentclass[reprint,longbibliography,aip]{revtex4-1}

\usepackage{graphicx} 
\usepackage{amsmath}
\usepackage{ amssymb }

\usepackage{color}

\usepackage{float} 
\usepackage{wrapfig} 

\usepackage[makeroom]{cancel} 
\usepackage{comment}

\usepackage{lipsum} 

\usepackage{outlines}

\newcommand{\beq}{\begin{equation}}
\newcommand{\eeq}{\end{equation}}
\newcommand{\vv}{\mathbf{v}}

\newcommand{\bvec}{\begin{pmatrix}}
\newcommand{\evec}{\end{pmatrix}}
\newcommand{\lp}{\left(}
\newcommand{\rp}{\right)}
\newcommand{\pa}[2]{\frac{\partial #1}{\partial #2}}

\newcommand{\paf}[2]{\partial #1 / \partial #2}
\newcommand{\llangle}{\left \langle}
\newcommand{\rrangle}{\right \rangle}

\newcommand{\ve}[1]{\mathbf{#1}}

\newcommand{\tn}{\tilde{n}}

\newcommand{\tphi}{\tilde{\phi}}
\newcommand{\tE}{\tilde{E}}

\renewcommand{\Im}[1]{\text{Im} #1}
\renewcommand{\Re}[1]{\text{Re} #1}

\newcommand{\tB}{\tilde{B}}
\newcommand{\tu}{\tilde{u}}

\AtBeginDocument{%
	\newwrite\bibnotes
	\def\bibnotesext{Notes.bib}
	\immediate\openout\bibnotes=\jobname\bibnotesext
	\immediate\write\bibnotes{@CONTROL{REVTEX41Control}}
	\immediate\write\bibnotes{@CONTROL{%
			aip41Control,author="08",editor="1",pages="1",title="0",year="1"}}
	\if@filesw
	\immediate\write\@auxout{\string\citation{aip41Control}}%
	\fi
}%

\begin{document}



\title{Wave-Driven Torques to Drive Current and Rotation}

\author{Ian E. Ochs}
\email{iochs@princeton.edu}
\affiliation{Department of Astrophysical Sciences, Princeton University, Princeton, New Jersey 08540, USA}
\author{Nathaniel J. Fisch}
\affiliation{Department of Astrophysical Sciences, Princeton University, Princeton, New Jersey 08540, USA}


\date{\today}

\begin{abstract}
	
In the classic Landau damping initial value problem, where a planar electrostatic wave transfers energy and momentum to resonant electrons, a recoil reaction occurs in the nonresonant particles to ensure momentum conservation. To explain how net current can be driven in spite of this conservation, the literature often appeals to mechanisms that transfer this nonresonant recoil momentum to ions, which carry negligible current. However, this explanation does not allow the transport of net charge across magnetic field lines, precluding $E\times B$ rotation drive. Here, we show that in steady state, this picture of current drive is incomplete. Using a simple Fresnel model of the plasma, we show that for lower hybrid waves, the electromagnetic energy flux (Poynting vector) and momentum flux (Maxwell stress tensor) associated with the evanescent vacuum wave, become the Minkowski energy flux and momentum flux in the plasma, and are ultimately transferred to resonant particles. Thus, the torque delivered to the resonant particles is ultimately supplied by the electromagnetic torque from the antenna, allowing the nonresonant recoil response to vanish and rotation to be driven. We present a warm fluid model that explains how this momentum conservation works out locally, via a Reynolds stress that does not appear in the 1D initial value problem. This model is the simplest that can capture both the nonresonant recoil reaction in the initial-value problem, and the absence of a nonresonant recoil in the steady-state boundary value problem, thus forbidding rotation drive in the former while allowing it in the latter.
	
\end{abstract}

\maketitle

\section{Introduction} 

Consider an electrostatic wave propagating through an unmagnetized plasma.
For certain particles in the plasma, the particle velocity happens to match the phase velocity of the wave.
These particles, which see the same wave phase for an extended time, are said to be ``Landau resonant,'' and can exchange energy and momentum efficiently with the wave.
Whether the wave adds or subtracts energy to each particle is essentially random, depending on the wave phase, and so averaged over an ensemble of resonant particles the wave interaction results in diffusion of the particles in momentum and energy.

To this simple system, add a background magnetic field along an axis $\hat{z}$, chosen so that the frequency of particle gyration $\Omega$ around this field is much lower that the wave frequency $\omega$.
If we choose our wavevector $\ve{k} \parallel \hat{z}$, then the resonant particles are pushed along the magnetic field, resulting in a current.
This setup is the basis for much of the wave-based current drive used in tokamaks \cite{Fisch1978}.
Alternatively, if we choose our wavevector $\ve{k} \perp \hat{z}$, then the gyrocenters of the resonant particles, which are determined by the particle canonical momentum, diffuse along the third direction $\ve{k} \times \hat{b}$.
This gyrocenter diffusion is the basis for alpha channeling \cite{Fisch1992,fisch1992current}.

While we will focus here on alpha channeling via lower hybrid waves \cite{Fisch1992,fisch1992current,Heikkinen1996,ochs2015coupling,ochs2015alpha}, one can also make use of waves in the ion-cyclotron range of frequencies \cite{Valeo1994,fisch1995ibw,Fisch1995a,Heikkinen1995,Marchenko1998,Kuley2011,Sasaki2011,Chen2016a,Gorelenkov2016,Cook2017,Castaldo2019,Romanelli2020}, and can further optimize the effect by combining multiple waves \cite{Herrmann1997,Cianfrani2018,Cianfrani2019,White2021}.
For any of these schemes, the basic idea is to set up the gyrocenter diffusion path so that hot, fusion-born alpha particles at the plasma center are cooled as they diffuse out of the plasma, thus transferring their energy into the waves, which can then be used either to drive current or heat fuel ions.

One of the most intriguing proposed applications of alpha channeling is to drive rotation in axially-magnetized plasmas \cite{fetterman2008alpha,fetterman2012wave}.
The basic idea is to manipulate particles of a certain type to on average diffuse from a source at the plasma center to a sink at the plasma edge, so that the net charge of these particles is extracted from the plasma.
This creates a radial electric field in the plasma, which combines with the axial magnetic field to drive rotation.
Such rotation, if sheared, can suppress turbulent transport \cite{maggs2007transition,Burrell2020} and stabilize the plasma \cite{Shumlak1995,Huang2001}.
In addition, this alpha channeling paradigm alters the energy flow in the plasma, as the waves can transfer energy from the alpha particles directly into the plasma rotation. 
Interestingly, in certain regimes, the viscous dissipation of this plasma rotation can disproportionately heat the ions, allowing the plasma to achieve a natural hot-ion mode without the use of direct ion heating \cite{Kolmes2021a}.

However, the success of these schemes, for both current and rotation drive, depends on the response of the nonresonant particles.
Although each nonresonant particle interacts only very weakly with the wave, there are many more nonresonant than resonant particles, so the many weak responses can add up.

The importance of this nonresonant response is particularly notable in a classic plasma physics problem relevant to current drive: the \emph{bump-on-tail instability} \cite{kralltrivelpiece,davidson1973methods,stix1992waves}.
In this purely one-dimensional problem, a high-frequency electrostatic wave interacts with electrons in an unmagnetized plasma, and the kinetic distribution is set up in such a way that energy is transferred from the resonant electrons into the waves, and the waves grow.
In the process, the resonant electrons lose momentum as well.
However, the electrostatic field of the wave contains no average momentum, and so the momentum lost from the resonant particles ends up in the nonresonant particles.
For a wave that interacts only with electrons, this means that no net current is driven after all.

\begin{figure}[tp]
	\center
	\includegraphics[width=0.8\linewidth]{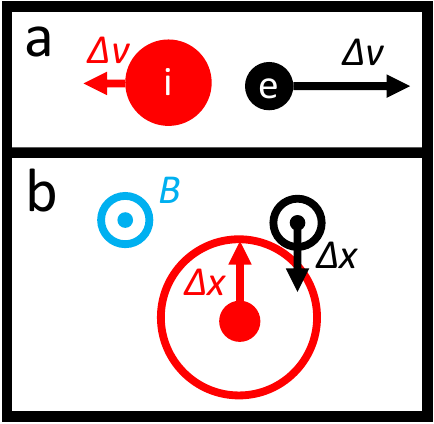}
	\caption{
		Consequence of the nonresonant recoil in current vs.~rotation drive.
		(a) For current drive, the nonresonant recoil momentum can be put into the ions, which contribute negligibly to the current, allowing for current drive despite the recoil.
		(b) For cross-field charge transport to drive $\ve{E} \times \ve{B}$ rotation, even if the nonresonant recoil is put into ions, the gyrocenter shifts of ions and electrons will cancel, so that no net charge moves across field lines.
	}
	\label{fig:currentVsRotation}
\end{figure}

Despite this theoretical result, experiments have long demonstrated that currents are, in fact, driven by waves in tokamaks \cite{Yamamoto1980,Wong1980,Kojima1981,Bernabei1982,Porkolab1984,Ekedahl1998}.
Often, this is attributed to some mechanism to put the nonresonant momentum into the ions, which are much heavier than the electrons and thus contribute negligibly to the current (Fig.~\ref{fig:currentVsRotation}a).
This offloading of nonresonant momentum into the ions can be accomplished directly by an appropriately chosen wave \cite{kato1980electrostatic,Ochs2020,Ochs2020a}.
Alternatively, because wave frequencies are typically chosen so as to drive resonant currents in the high-energy tail electrons, the inverse energy dependence of the Coulomb cross section causes bulk thermal electrons to transfer their momentum to ions much more quickly than resonant electrons \cite{bellan2008fundamentals,stix1992waves,fisch1978currentDrive,fisch1980creating}.

If these explanations for how current drive works were correct, then rotation drive via alpha channeling would be impossible.
In a uniformly magnetized plasma with no electric field, the gyrocenter position of a particle is intrinsically linked to its momentum. 
As a result, if two particles exchange momentum, their gyrocenters will move in such a way that no net charge moves (Fig.~\ref{fig:currentVsRotation}b).
This link between charge transport and momentum conservation is why classical transport \cite{braginskii1965transport} is to lowest order ambipolar \cite{helander2005collisional,Rozhansky2008,ochs2018favorable,kolmes2019radial}.
It is also responsible for the cancellation \cite{Chankin1994,Chankin1996,Stoltzfus-Dueck2019} of proposed radial currents \cite{Rozhansky1994} in the study of intrinsic rotation in tokamaks.
Thus, if equal and opposite forces are applied to the resonant and nonresonant particles, then regardless of which species the nonresonant force is applied to, no net charge can be transported: the resonant charge transport will simply be cancelled by a nonresonant charge transport, making rotation drive impossible.
Indeed, just such a cancellation was recently shown to exist for this 1D initial value problem in a magnetized plasma \cite{ochs2021nonresonant}.

Fortunately, as we will discuss in this paper, the above explanation for current drive is incorrect.
There is a fundamental difference between how momentum conservation works in a multi-dimensional boundary-value problem (BVP) compared to a one-dimensional initial-value problem (IVP). As a result, in many cases of interest, the cancelling current or charge transport that appears in the IVP is absent in the BVP, allowing both current and rotation drive immediately.

The precise way in which the boundary value problem differs from the initial value problem has been the focus of some conflicting explanations.
For instance, it has been recognized that, in the BVP, an electrostatic wave is really only quasi-electrostatic, carrying a magnetic field which allows energy flow into the plasma (see e.g. Sec.~16.7 of \cite{stix1992waves}).
It has also been recognized that the off-diagonal component of the generalized stress tensor in the presence of a wave plays a crucial role in driving rotation. 
The relative importance of these two sources has not been extensively explored.

Much of the previous work in the area of wave-driven rotation has focused on low-frequency electrostatic turbulence \cite{Berk1983,Diamond1991,Diamond2008,Krommes2013}.
Often in this work, key assumptions appropriate to low-frequency turbulence were made, which make the analysis less suitable for our current and rotation drive mechanisms.
For instance, dissipation due to resonances was often neglected or assumed to be in detailed balance, whereas such dissipation provides the basis for current and charge transport in our mechanisms.
In addition, a key part of some theorems \cite{Diamond1991} involved replacing the radial velocity with the wave-induced $E \times B$ velocity, an approximation valid only for low-frequency waves.

A second category of prior work in this area focused on hot, kinetic, magnetized plasmas  \cite{Berry1999,Jaeger2000,Myra2000,Myra2002,Myra2004}.
However, these models focused exclusively on the steady-state boundary-value problem, rather than incorporating the initial value problem as well.
In addition, the complexity of the kinetic mathematics has a couple drawbacks.
First, it makes the theory tricky to generalize to more complex plasmas, such as those with shear flow.
Second, it obscures possible issues with the calculation, such as a failure for some of the theories to agree with the cold-fluid ponderomotive force in the appropriate limit, which have taken years to emerge \cite{Chen2013}.

A third area of prior work \cite{Lee2012,guan2013plasma,guan2013toroidal}, focused on deficiencies of the Kennel-Engelmann quasilinear theory \cite{kennel1966magnetizedQL} in describing perpendicular momentum effects in hot-plasma current drive problems, ignored the nonresonant particles entirely.
However, these papers exposed important simplifying features of the resonant particle behavior, showing how the resonant particle momentum absorption was consistent with the absorption of Minkowski momentum from geometrical optics \cite{dodin2012axiomatic}.

In this paper, we aim to put forth the simplest possible model that captures the behavior in both the initial and boundary value problem, and exposes the core differences that allow for momentum injection in the boundary value problem but not the initial value problem.
To this end, we begin in Sec.~\ref{sec:EMTs} with an overview of the different types of energy and momentum commonly encountered in plasma wave problems: the electromagnetic energy-momentum, the particle energy momentum, and the Minkowski energy-momentum.
These energy and momenta, which can be grouped into energy-momentum tensors, form a couple different closed systems.
We use these concepts in Sec.~\ref{sec:1DIVP} to review how in the 1D initial value problem, no net momentum is transferred into the particles.

In Sec.~\ref{sec:injectionModel}, we introduce the overall framework for wave injection we use throughout the paper.
The subsequent two sections use this framework to show how momentum conservation works out globally, and locally in the area of wave damping, for lower hybrid current and rotation drive.

In Sec.~\ref{sec:FresnelModel}, we adopt a simple Fresnel model for wave injection into the plasma, that is broadly consistent with the theory \cite{Brambilla1976,Brambilla1979} of low-frequency ($\omega \ll \Omega_e$) wave launching into a plasma.
In this model, an evanescent wave in a homogeneous vacuum region converts into a traveling slow wave in a bordering low-density plasma region.
We show that, in steady state, the flux of \emph{electromagnetic} energy and non-radial momentum through the evanescent vacuum region, as given by the Poynting flux and Maxwell stress tensor, are identical to the flux of the \emph{Minkowski} energy and non-radial momentum in the plasma region.
This result stands in contrast to the initial value problem, where the electromagnetic energy and momentum are in general different from the Minkowski energy and momentum \cite{Ochs2020}.
As a consequence, we show that in the boundary-value problem, all the energy and momentum that ends up in the resonant particles is ultimately supplied by the fields near the waveguide, suggesting that the nonresonant response vanishes.
This fact establishes global momentum conservation.

In Sec.~\ref{sec:WarmFluidModel}, we focus on the region of wave damping, after the wave has mode-converted into an electrostatic wave.
Here, we show how the absence of the nonresonant response can be understood to be consistent with momentum conservation, arising partially (as in earlier works) from a wave-induced off-diagonal component of the stress tensor.
We show that the minimal model required to understand the behavior is the warm-fluid model, which is related to the fact that it gives a wave with non-vanishing group velocity.
Using the warm-fluid model, we recover the behavior in both the initial value problem, where no momentum is injected into the plasma, and in the boundary-value problem, where momentum is injected exclusively into the resonant particles.
This result establishes the local momentum conservation of the theory.

\section{Forms of energy and momentum} \label{sec:EMTs}

Before we delve into the specific scenario we will study for wave-driven rotation, we review the types of energy and momentum that will be important throughout the paper.

Energy, momentum, and their associated fluxes are often grouped together into a single object $\ve{T}$ known as an energy-momentum tensor (EMT):
\begin{align}
	\ve{T} &= \bvec W & \ve{S}/c \\ c \ve{p} & \boldsymbol{\Pi} \evec.
\end{align}
Here, $W$ is the energy, $\ve{p}$ is the momentum, $\ve{S}$ is the energy flux, and $\boldsymbol{\Pi}$ is the momentum flux, also called the stress.

In a closed system in space, energy and momentum conservation are expressed by the vanishing of 4-divergence of the energy-momentum tensor, i.e:
\begin{align}
	\nabla_\nu T^{\mu \nu} = 0,
\end{align}
where $\nabla_\nu$ is the covariant derivative \cite{carroll2003spacetime}, and we use Einstein summation notation.
For flat spacetime and Cartesian coordinates, this becomes:
\begin{align}
	\frac{1}{c} \pa{}{t} T^{\mu 0} + \pa{}{x_i} T^{\mu i} &= 0. \label{eq:stressTensorDiv0}
\end{align}
Energy conservation corresponds to the $\mu = 0$ portion of this equation, and momentum conservation to the $\mu =  1-3$ components.

Eq.~(\ref{eq:stressTensorDiv0}) applies to the stress tensor which encompasses all forms of energy and momentum in the system.
Often, it is useful to separate out different subsystems.
For instance, in a plasma, we can separate out the electromagnetic and particle subsystems:
\begin{align}
	\ve{T} = \ve{T}_{EM} + \ve{T}_{P}. \label{eq:consEMP}
\end{align}
Then, Eq.~(\ref{eq:stressTensorDiv0}) applies to this sum.
The components of the electromagnetic subsystem are given by:
\begin{align}
	W_{EM} &= \frac{E^2 + B^2}{8\pi} \label{eq:wEM}\\
	S_{EM}^i &= \frac{c}{4\pi }\epsilon^{ijk} E_j B_k \label{eq:sEM}\\
	p_{EM}^i &= \frac{S^i}{c^2} \label{eq:pEM}\\
	\Pi_{EM}^{ij} &= -\frac{1}{4\pi} \lp E^i E^j - \frac{1}{2} \delta^{ij} E^2 +  B^i B^j - \frac{1}{2} \delta^{ij} B^2 \rp \label{eq:piEM},
\end{align}
where $E$ is the electric field and $B$ is the magnetic field.
Meanwhile the components of the particle subsystem $P$ are given by \cite{Blandford2017}:
\begin{align}
	T^{\mu \nu}_{P} &= \int f_p(t,\ve{x},\ve{p}) p^\mu p^\nu \frac{d \ve{p}}{p^0}\\
	p^0 &= U_P = \sqrt{m^2 c^4 + |\ve{p}|^2 c^2},
\end{align}
where $p$ here is the relativistic 4-momentum $(U_P,\ve{p}_P)$, so that $p^0 = U_P$ and $p^i = m v^i/(1-v^2/c^2)$, and $f_p$ is the distribution function in momentum space.
In the sub-relativistic limit $v \ll c$, the components of the tensor become:
\begin{align}
	W_{P} &= \int  \lp m c^2 + \frac{1}{2} m v^2 \rp f(t,\ve{x},\ve{v}) d \ve{v} \label{eq:wP}\\
	S_{P}^i &= \int \lp m c^2 + \frac{1}{2} m v^2 \rp v^{i}  f (t,\ve{x},\ve{v}) d \ve{v} \label{eq:sP}\\
	p_{P}^i &= \int  m v^{i} f (t,\ve{x},\ve{v}) d \ve{v} \label{eq:pP}\\
	\Pi_{P}^{ij} &= \int  m v^{i} v^j f (t,\ve{x},\ve{v}) d \ve{v} , \label{eq:piP}
\end{align}

In addition to these familiar and physically intuitive forms of energy and momentum, there is another energy-momentum tensor we can form for waves in a plasma, known as the Minkowski energy-momentum tensor \cite{dodin2012axiomatic}.
The components of the Minkowski EMT are expressed in terms of the wave action $\mathcal{I}$ and group velocity $\vv_g$:
\begin{align}
	W_M &= \omega_r \mathcal{I} \label{eq:wMinkowski}\\
	S^i_M &= W_M v_g^i \label{eq:sMinkowski}\\
	p^i_M &= k_r^i \mathcal{I} \label{eq:pMinkowski} \\
	\Pi^{ij}_M &= p^i_M v_g^j. \label{eq:piMinkowski}
\end{align}
Consider a quasi-monochromatic electromagnetic wave with complex amplitude $\ve{\tE}$, such that the physical wave is given by:
\begin{align}
	\ve{E} = \Re \lp \ve{\tE} e^{i\ve{k} \cdot \ve{x} - i\omega t} \rp \label{eq:EfromEc}
\end{align}
For such a wave, when the magnetic susceptibility $\mu = 1$, the action is given by \cite{dodin2012axiomatic}:
\begin{align}
	\mathcal{I} &= \frac{1}{16 \pi \omega^2} \tE^{i*} \pa{}{\omega_r} \lp \omega^2 \epsilon_{H,ij} \rp \tE^j. \label{eq:actionElectromagnetic}
\end{align}
Here, $\epsilon_{H,ij}$ is the Hermitian part of the dielectric tensor \cite{stix1992waves}.
The group velocity is obtained from the part $D_r$ of the dispersion function $D$ that is real when $D$ is evaluated at real $\omega, k$:
\begin{align}
	v_g^i &= -\frac{\paf{D_r}{k_{ri}}}{\paf{D_r}{\omega_r}}. \label{eq:groupVelocity}
\end{align}
If the polarization is known, the dispersion function can be written in terms of the polarization vector $\ve{\tilde{e}}_c \equiv \ve{\tE} / |\ve{\tE}|$ \cite{dodin2012axiomatic}:
\begin{align}
	D &= \left[\ve{\tilde{e}}_c^{*} \cdot \boldsymbol{\epsilon} \cdot \ve{\tilde{e}}_c - \frac{c^2}{\omega^2} \lp \ve{k} \times \ve{\tilde{e}}^{*}_c \rp \cdot \lp \ve{k} \times \ve{\tilde{e}}_c \rp \right]. \label{eq:dispersionPolarizationForm}
\end{align}

There are a couple limitations of the Minkowski EMT that must be noted.
First, the Minkowski EMT is only well-defined in areas where the wavepacket is \emph{eikonal}, i.e. $|k_i| \ll |k_r|$ and $|\omega_i| \ll |\omega_r|$.
Thus, it is not well-defined in regions where the wave is evanescent.

Second, there is no relevant EMT that combines with the Minkowski EMT in such a way that Eq.~(\ref{eq:stressTensorDiv0}) is satisfied for the combination of systems. 
In other words, the Minkowski EMT does not form a physically-relevant subsystem of the closed plasma-wave system.

Nevertheless, the Minkowski energy-momentum tensor is extremely useful for ray tracing calculations and calculating the evolution of wavepackets.
Furthermore, since it incorporates the energy of the oscillating particles, it provides an intuitive notion of ``total'' wave energy.

In regions where geometrical optics applies and the wave remains eikonal, the evolution of the Minkowski EMT follows from the conservation of the wave action \cite{dodin2012axiomatic} [should put some other references here]:
\begin{align}
	\pa{\mathcal{I}}{t} + &\nabla \cdot \lp \vv_g \mathcal{I} \rp =  -\Gamma \mathcal{I}. \label{eq:actionConservation}\\
	&\Gamma \equiv \frac{2 D_i}{\paf{D_r}{\omega_r}} \label{eq:actionDissipationGamma}
\end{align}
The right hand side of Eq.~(\ref{eq:actionConservation}) represents dissipation on resonant particles.
Thus, in areas where $D_r$ is purely real---i.e. in areas without resonant particles interactions---the action is perfectly conserved, and advected at the group velocity.

This action conservation principle is useful even in wave problems where the eikonal approximation breaks down in a local region, known as a caustic. 
Such caustics occur in regions of wave reflection, tunneling, or mode conversion.
In caustic regions, the wave action on rays flowing into the caustic reappears on rays flowing out of the caustic, and action is still conserved [should double check that this is always 100\% true] \cite{Tracy2014}.

Though the Minkowski EMT does not generally combine with anything to form a closed subsystem of a total EMT, it can be shown from Eq.~(\ref{eq:actionConservation}) that in the special case when $k_r$ and $\omega_r$ are constant, all energy and momentum that are lost from the Minkowski EMT show up in the \emph{resonant particle} EMT \cite{Diamond2008,guan2013plasma}.
I.e., if there are no other forces on the resonant particles $R$, then
\begin{align}
	\ve{T} = \ve{T}_M + \ve{T}_{R} \label{eq:consMRP}
\end{align}
forms a closed system.
Here, the components of the resonant particle EMT $T_R$ are similar to the components of the total particle EMT $T_P$, except the integrations are performed only over the resonant parts of the particle distribution, i.e. over the region where $\ve{k} \cdot \vv \approx \omega_r$ for a Landau resonance. 
We will make use of this extremely useful property throughout the paper.

For a propagating light wave in a vacuum, the electromagnetic EMT and Minkowski EMT coincide \cite{dodin2012axiomatic}.
However, in general the two EMTs can be completely different, as we review in the next section.

\section{1D initial value problem example} \label{sec:1DIVP}

As an example of the stark difference between the Minkowski EMT and the electromagnetic EMT, and of the ways in which our various conservation laws are useful, consider the classic problem of Landau damping of electron Langmuir waves in an unmagnetized plasma \cite{kralltrivelpiece,davidson1973methods,stix1992waves}.
For this problem, the ions barely interact with the wave and can be ignored.
We proceed quickly, since similar discussions can be found in Refs.~\cite{dodin2012axiomatic,Ochs2020,Ochs2020a}.

The unmagnetized, hot-plasma susceptibility tensor is \cite{kralltrivelpiece,stix1992waves}:
\begin{align}
	\epsilon_{ij} &= \delta_{ij} \left[ 1 - \frac{\omega_{pe}^2}{k^2} \int_\mathcal{L} d \ve{v} \frac{\ve{k} \cdot \paf{f_{e0}}{\ve{v}}}{\ve{k}\cdot \ve{v} - \omega} \right],
\end{align}
where $\delta_{ij}$ is the Kronecker delta function, $f_{e0}$ is the electron distribution function in velocity space (normalized to one), $\omega_{pe} = (4\pi e^2 n_e / m_e)^{1/2}$ is the electron plasma frequency, and $e$, $m_e$, and $n_e$, are the electron charge, mass, and average number density, respectively. 
The integral is performed along the Landau contour, wrapping around singularities as necessary to analytically continue the contour from $\omega_i \gg 1$.

For this 1D problem, $k$ is purely real, and we will take $\ve{k} \parallel \hat{y}$.
Then, the integral can be trivially integrated over $v_x$ and $v_z$, and we can reduce the integral to a 1D integral with $f_{e0}$ replaced by $g_{e0} = \int dv_x dv_z f_{e0}$.
Then, taking the standard expansions $|\omega_i / \omega_r| \ll 1$ and  $|\ve{k} \cdot\ve{v} / \omega_r| \ll 1$, we find to lowest order the Hermitian and anti-Hermitian susceptibilities:
\begin{align}
	\epsilon_{H,ij} &= \delta_{ij} \left[1 - \frac{\omega_{pe}^2}{\omega^2} \right]\\
	\epsilon_{A,ij} &= - \delta_{ij} i \pi \frac{\omega_{ps}^2}{k^2} \pa{g_{e0}}{v_y} \biggr|_{\omega_r/k}.
\end{align} 

Because we are examining an electrostatic wave, we have $\ve{\tE} \parallel \ve{k}$.
Without loss of generality, we can take $\ve{\tE} \in \mathbb{R}$ (this merely sets the wave phase), so that $\ve{\tilde{e}}_c = \hat{y}$.
Then, our wave action is given from Eq.~(\ref{eq:actionElectromagnetic}) by:
\begin{align}
	\mathcal{I} &= \frac{|\tE^y_c|^2}{8 \pi \omega},
\end{align}
and our dispersion function is given from Eq.~(\ref{eq:dispersionPolarizationForm}) by:
\begin{align}
	D &= 1 - \frac{\omega_{pe}^2}{\omega^2} - i \pi \frac{\omega_{pe}^2}{k^2} \pa{g_{e0}}{v_y} \biggr|_{\omega_r/k},
\end{align}
from which
\begin{align}
	\omega_r^2 &= \omega_{pe}^2\\
	\Gamma &= -\pi \frac{\omega_r^3}{k^2} \pa{g_{e0}}{v_y} \biggr|_{\omega_r/k}
\end{align}

Because this problem is homogeneous and 1D, the action conservation equation~(\ref{eq:actionConservation}) takes the simple form:
\begin{align}
	\pa{\mathcal{I}}{t} &= -\Gamma \mathcal{I}.
\end{align}
Thus, when $\paf{g_{e0}}{v_y} < 0$, the action will decay, consistent with Landau damping.
As long as the wave amplitude is small enough for the quasilinear theory to be valid, the action will asymptotically approach 0.

For this problem, we can easily calculate the components of both the Minkowski and electromagnetic EMTs. 
Since this is an initial value problem with no spatial variation, we ignore the flux terms $\ve{S}$ and $\boldsymbol{\Pi}$. 
Our Minkowski energy and momentum are given by:
\begin{align}
	W_{M} = \frac{|\tE^y_c|^2}{8 \pi} \qquad p^y_{M} = \frac{k_y}{\omega_r} W_M.
\end{align}
Meanwhile, our electromagnetic energy and momentum are given by:
\begin{align}
	W_{EM} = \frac{|\tE^y_c|^2}{16 \pi} \qquad p^y_{EM} = 0.
\end{align}
Thus, the Minkowski energy of the wave is double the electromagnetic energy of the wave, and the wave has Minkowski momentum, but no electromagnetic momentum.

To see how this plays out physically as the wave damps, we can make use of our closed systems and conservation relation \ref{eq:stressTensorDiv0}.
From the closed EMT in Eq.~(\ref{eq:consMRP}), we find:
\begin{align}
	\Delta W_{RP} = W_{M0} \qquad \Delta p^y_{RP} = p^y_{M0},
\end{align}
while from the closed EMT in Eq.~(\ref{eq:consEMP}) we find:
\begin{align}
	\Delta W_{P} = W_{EM0} \qquad \Delta p^y_{P} = p^y_{EM0},
\end{align}
where the $\Delta$ indicates the change in the quantity once the wave has completely damped, and the subscript 0 represents the initial value.
Together, these imply a relation for the nonresonant particles $NP$:
\begin{align}
	\Delta W_{NP} &= \Delta (W_P - W_{RP}) = - \Delta W_{RP} / 2\\ 
	\Delta p^y_{NP} &= \Delta (p^y_{P} - p^y_{RP})   = - \Delta p^y_{RP}. 
\end{align}
Thus, as the wave damps, all the Minkowski energy and momentum in the wave end up as physical energy and momentum in the resonant particles.
To be consistent with overall energy and momentum conservation of the electromagnetic-particle closed system, the nonresonant particle must lose energy, and gain momentum equal and opposite to the resonant particle momentum.
The nonresonant momentum shift thus cancels out the current driven in the resonant particles.

The loss of energy from the nonresonant particles can be understood as the loss of ``sloshing'' motion associated with the wave, rather than some sort of thermal cooling.
The resulting ``negative diffusion'' in energy space, which also arises in more detailed quasilinear calculations, was a source of consternation in the plasma waves community for some time, until it was shown to be consistent with energy and momentum conservation in this way \cite{Kaufman1972a}.

This simple example of Landau damping of a uniform plane electron Langmuir wave in a homogeneous unmagnetized plasma demonstrates both the power and limitations of the Minkowski energy-momentum, and the dangers of considering only the resonant particles when evaluating effects such as current drive and momentum damping.
Understanding the relationships between the various forms of momentum, as well as the behavior of the nonresonant particles, is thus key to understanding rotation and current drive.

\begin{figure*}[tp]
	\center
	\includegraphics[width=0.9\linewidth]{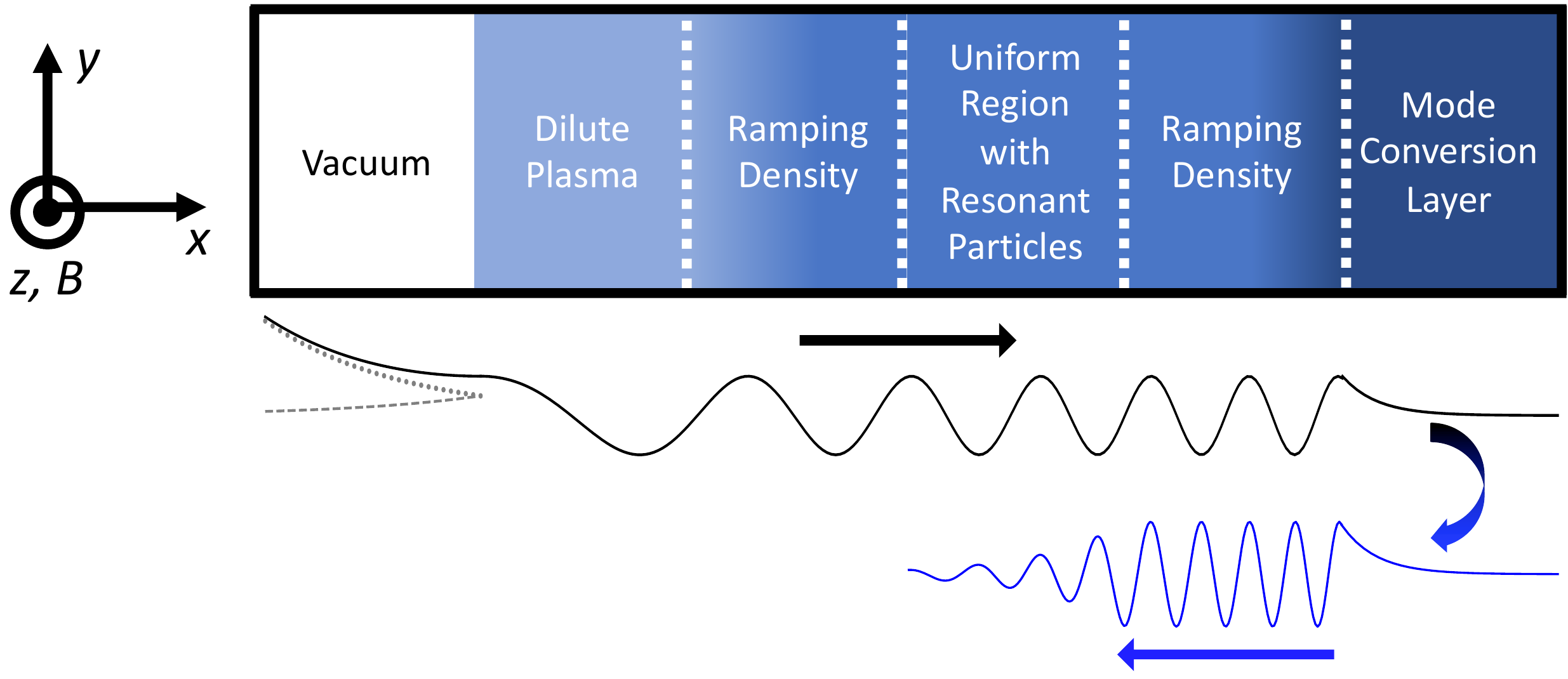}
	\caption{
		Coordinate system and wave injection model used throughout the paper.
		The coordinates $x$, $y$, and $z$ correspond to radial, azimuthal, and axial / toroidal directions, respectively.
		Darker color represents higher density.
		A vacuum evanescent wave (black), consisting of a forward-decaying ``incident'' wave (dotted gray) and backward-decaying ``reflected'' wave (dashed gray), converts to a ``transmitted'' slow wave at the plasma / vacuum interface.
		This slow wave then continues to propagate inward as the plasma density increases until it hits the lower hybrid resonance, where it mode converts into an electrostatic lower hybrid wave (blue).
		Wave action is conserved during the in-plasma propagation and mode conversion.
		This wave then propagates back to a uniform region containing resonant particles, where it damps.
	}
	\label{fig:waveInjectionModel}
\end{figure*}

\section{A simple model of wave injection} \label{sec:injectionModel}

We now turn our attention to the boundary value problem.
We design our overall model to make maximal use of the conservation properties of our system, while avoiding calculational complexity whereever possible.

The coordinate system and model of wave injection we use throughout the paper are shown in Fig.~\ref{fig:waveInjectionModel}.
We work in a slab geometry, with all gradients along the ``radial coordinate'' $x$ and a magnetic field along the ``axial'' or ``toroidal'' coordinate $z$, with $y$ taking the place of the ``poloidal'' coordinate.

Our simple model is motivated by the coupling of waveguides for lower hybrid current drive \cite{Brambilla1976,Brambilla1979}.
For such waves, an evanescent wave in a vacuum region converts into a plasma slow wave (also known as the extraordinary wave, or X-mode) as the plasma density ramps up past the point where $|\omega_{pe}| > |\omega|$.
To avoid calculational complexity associated with the slowly ramping density, which often necessitates numerical full-wave calculations, we instead consider a Fresnel-type model, where an evanescent wave in a homogeneous vacuum region converts into a propagating slow wave at a sharp boundary to a dilute plasma region.
This reduces the coupling calculation to a boundary-matching condition, dramatically simplifying the mathematics.

In the physics of lower hybrid coupling, assuming we have chosen a wave with $k_z^2 c^2 / \omega^2 > 1 + \omega_{pe}^2/\Omega_{ce}^2$, as the plasma density ramps up, the wave continues to propagate until it hits the lower hybrid resonance layer, where $|\omega| = |\omega_{LH}| = (\omega_{pi}^{-2} + |\Omega_e \Omega_i|^{-1})^{-1/2}$, with $\Omega_s = q_s B_0 / m_s c$ the cyclotron frequency.
At this point, the wave mode-converts into an outward-propagating lower hybrid wave \cite{Brambilla1976,Stix1965}.
We incorporate this process in our model by having a region of gradually ramping density, where we will make use of the action conservation both in the region where geometric optics applies, and in balancing action going into and out of the mode conversion layer. 

It is this mode-converted lower hybrid wave that interacts with the resonant particles strongly.
In order to simplify our analysis of the resonant particles, we will assume that resonant particles only exist in a uniform region in the middle the density ramping region.
This will allow us to use our closed system from Eq.~(\ref{eq:consMRP}) easily, though it is not actually essential to our final result.
We assume that there are enough resonant particles to completely deplete the energy in the wave, so that no wave energy propagates back out of the plasma after entering.

What we will show, in the next sections, is that for this system in steady state, the $x$-directed fluxes of electromagnetic energy $S^x_{EM}$ and $y$ and $z$ momentum $\Pi^{yx}_{EM}$ and $\Pi^{zx}_{EM}$ for the evanescent wave in the vacuum region, are equal to the $x$-directed fluxes of Minkowski energy $S^x_{M}$ and $y$ and $z$ momentum $\Pi^{yx}_{M}$ and $\Pi^{zx}_{M}$ for the propagating slow wave in the dilute plasma region.
Via action conservation, these are the same energy and momentum that eventually end up in the resonant particles.
In other words, the energy and momentum that are transferred to the resonant particles is supplied by electromagnetic energy and momentum through the waveguide-plasma gap.

The correspondence between the electromagnetic momentum and the momentum that ends up in the resonant particles strongly suggests that the nonresonant momentum should disappear in the purely boundary-value problem.
However, there is still a possibility that the wave would induce rearrangement of momentum within the plasma, perhaps locally canceling the resonant momentum but transferring an equivalent amount of momentum elsewhere.
Thus, in the subsequent sections, we will explicitly calculate the momentum balance within the resonant damping region using a warm fluid theory for the particles, and show how the absence of a nonresonant response is consistent with momentum conservation.
This warm fluid theory is capable of explaining both the initial value problem and the boundary value problem, demystifying questions of momentum conservation in current and rotation drive.

\section{Vacuum electromagnetic energy and momentum flux end up in resonant particles} \label{sec:FresnelModel}

\subsection{Vacuum: energy and momentum flux relation}

Our problem has two symmetry directions: $\hat{y}$ and $\hat{z}$. 
In this section, we will show that for any electromagnetic wave in a vacuum, either propagating or evanescent along the non-symmetry direction $\hat{x}$, the fluxes of electromagnetic energy and momentum are related by:
\begin{align}
	\Pi^{yx}_{EM} = \frac{k_y}{\omega} S^x \qquad \Pi^{zx}_{EM} = \frac{k_z}{\omega} S^x. \label{eq:vacuumEMEnergyMomentumFluxRelation}
\end{align}
Since this is the same relation as exists between the Minkowski energy flux $S^x_M$ and momentum flux $\Pi^{ix}_M$, this will mean that we only have to demonstrate the equivalence of the vacuum electromagnetic energy and Minkowski energy to show the equivalence of the symmetry-direction momentum fluxes as well.

We will begin by rotating our coordinate system about the $x$ axis to a new set of coordinates $(x,u,v)$, so that the wavevector lies in the $x$-$u$ plane.
Explicitly, we take $\hat{u} \parallel k_y \hat{y} + k_z \hat{z}$, and $\hat{v} = \hat{x} \times \hat{u}$.
Then, defining the refractive index $\ve{n} = \ve{k} c/\omega$, the dispersion relation is given from Fourier transforming Maxwell's equations as:
\begin{align}
	\bvec 1- n_u^2 & n_x n_u & 0\\ n_x n_u & 1-n_x^2 & 0\\ 0 & 0 & 1-n_x^2 -n_u^2\evec \bvec \tE^x \\ \tE^y \\ \tE^z \evec &= 0.
\end{align}
Taking the determinant of the left matrix leads to the single dispersion relation for electromagnetic waves in a vacuum:
\begin{align}
	D = 1 - n_x^2 - n_u^2 = 0.
\end{align}
Any such waves will be a linear combination of two possible polarizations: the $p$ polarization, given by:
\begin{align}
	\ve{\tE}_{c,p} &= n_u E_0 \hat{x} - n_x E_0 \hat{u},
\end{align}
and the $s$ polarization, given by:
\begin{align}
	\ve{\tE}_{c,s} &= E_0 \hat{v},
\end{align}
where $E_0$ is an arbitrary complex constant.
Note that no assumption has been made as to whether the components of $\ve{n}$ are real or complex.
From Faraday's Law, $\ve{\tB}_{c} = \ve{n} \times \ve{\tE}$, so the magnetic field components corresponding to the $p$ and $s$ polarizations are:
\begin{align}
	\ve{\tB}_{c,p} &= - (n_x^2 + n_u^2) E_0 \hat{v} = E_0 \hat{v}\\
	\ve{\tB}_{c,s} &= n_u E_0 \hat{x} - n_x E_0 \hat{u}
\end{align}

As we calculate the relation between the energy and momentum fluxes, we will focus only on the $p$ polarization.
The reason that we are able to do this is that we can easily show that the same relations hold if the wave is in $s$ polarization.
To see this, tote that the above relations imply that:
\begin{align}
	\ve{\tE}_{c,s} = -\ve{\tB}_{c,p} \qquad \ve{\tB}_{c,s} = \ve{\tE}_{c,p}.
\end{align}
These in turn imply that:
\begin{align}
	\ve{E}_{s} = -\ve{B}_{p} \qquad \quad  \ve{B}_{s} = \ve{E}_{p}.
\end{align}
Thus,
\begin{align}
	\ve{S}_{EM,s} &= \frac{\ve{E}_s \times \ve{B}_s}{4\pi} = \frac{(-\ve{B}_p) \times \ve{E}_p}{4\pi} = \frac{\ve{E}_p \times \ve{B}_p}{4\pi} \\
	&= \ve{S}_{EM,p}\\
	\boldsymbol{\Pi}_{EM,s} &= -\frac{1}{4\pi} \lp \ve{E}_s\ve{E}_s - \frac{1}{2} E_s^2 \ve{I} + \ve{B}_s\ve{B}_s - \frac{1}{2} B_s^2 \ve{I} \rp \\
	 &= -\frac{1}{4\pi} \lp \ve{B}_p\ve{B}_p - \frac{1}{2} B_p^2 \ve{I} + \ve{E}_p\ve{E}_p - \frac{1}{2} E_p^2 \ve{I} \rp\\
	&= \boldsymbol{\Pi}_{EM,p},
\end{align}
and it is sufficient to prove the relation only for the $p$ polarization.

From here, the proof is very short.
In general, there will be an ``incident'' wave $I$, with $\Im(n_{xI}) >0$, and a ``reflected'' wave $R$, with $\Im(n_{xR}) < 0$ (we use this terminology because it is familiar from Fresnel calculations in introductory electromagnetism \cite{Griffiths2017,Jackson1999}.
It is important to consider both these fields together, rather than independently, because it will turn out to be the interplay between the incident and reflected wave fields that allow power to flow through the vacuum.
The refractive indices of the incident and reflected waves satisfy $n_{xR} = -n_{xI}$ and $n_{uR} = n_{uI} \equiv n_u$.
Furthermore, as $u$ is a symmetry direction, $n_{u}$ is purely real.
From these two relations, Eq.~(\ref{eq:EfromEc}), and the definitions of $\ve{\tE}_{c,p}$ and $\ve{\tB}_{c,p}$, it follows that:
\begin{align}
	E_{p}^x &= \Re \left[\lp -n_{uI} \times E_{0I} e^{i\ve{k}_I \cdot \ve{x}-i\omega t}- n_{uR} \times E_{0I} e^{i\ve{k}_R \cdot \ve{x}-i\omega t}\rp \right]\\
	&=-n_{u} B_{p}^v.
\end{align}
We can now calculate the components of the stress tensor in the $(x,u,v)$ coordinate system.
Recalling that $n_u = k_u c/\omega$, we have:
\begin{align}
	\Pi^{ux}_{EM} &= -\frac{1}{4\pi} \llangle E^u E^x \rrangle = \frac{k_u}{\omega} \frac{c}{4\pi} \llangle E^u B^v \rrangle =  \frac{k_{u}}{\omega} S^x_{EM} \\
	\Pi^{vx}_{EM} &= 0.
\end{align}
Rotating our coordinate system to $(x,y,z)$ recovers the relation in Eq.~(\ref{eq:vacuumEMEnergyMomentumFluxRelation}).

With this proof in hand, we now only need to show the equivalence of $S^x_{EM}$ in the vacuum to $S^x_M$ in the plasma.
Once this is established, the equivalence of $\Pi^{ix}_{EM}$ and $\Pi^{ix}_M$ for $i \in \{y,z\}$ follows automatically from Eq.~(\ref{eq:vacuumEMEnergyMomentumFluxRelation}) and Eqs.~(\ref{eq:sMinkowski}-\ref{eq:piMinkowski}).

\subsection{Fresnel equations for vacuum-plasma transition}

Showing the equivalence of the electromagnetic and Minkowski energy fluxes requires relating the electric fields in the vacuum in plasma using boundary matching conditions at the interface, which we now turn to.

We are interested in calculating wave propagation through the vacuum, which we denote region $1$, and the dilute boundary region of the plasma, which we denote region $2$ (Fig.~\ref{fig:waveInjectionModel}).
In these regions, a cold plasma model is sufficient.
For the slow waves that eventually become lower hybrid waves, we have $|\omega| \ll |\Omega_e|$, and we choose the plasma density at the edge such that $|\omega| \lesssim |\omega_{pe}|$.
For such a wave, the $S$-$P$-$D$ susceptibility tensor of Stix \cite{stix1992waves} becomes:
\begin{align}
	\boldsymbol{\epsilon} &= \bvec 1 & 0 & 0\\ 0 & 1 & 0 \\ 0 & 0 & P\evec \qquad P \equiv 1-\frac{\omega_{pe}^2}{\omega^2} .
\end{align}
Note that in the vacuum, this susceptibility tensor still works; we simply have $P_1 = 1$.

From this susceptibilty tensor, the dispersion relation is given from:
\begin{align}
	(n_i n_j - n^2 \delta_{ij} + \epsilon_{ij} ) E^j = 0.
\end{align}
Taking the determinant gives two dispersion branches.
The slow wave branch that we are interested in is:
\begin{align}
	n_x^2 + n_y^2 + P n_z^2 = P.
\end{align}
Plugging this back to the dispersion relation matrix equation in gives the polarization:
\begin{align}
	\tE^x_c &= \frac{n_x}{n_\perp} E_0; \quad
	\tE^y_c = \frac{n_y}{n_\perp} E_0; \quad 
	\tE^z_c = -\frac{1}{P} \frac{n_\perp}{n_z} E_0,
\end{align}
where $n_\perp = \sqrt{n_x^2 + n_y^2}$, and we take the root that is either positive or positive imaginary.
This polarization applies to both the wave in the vacuum, where $P_1 = 1$, and in the plasma, where $P_2 < 0$.

The dispersion relation can be put in the form:
\begin{align}
	n_\perp^2 &= P(1 - n_z^2).
\end{align}
Thus, in order for the plasma wave to propagate ($n_\perp^2 > 0$) in the plasma, where $P_2 < 0$, we must have $n_z^2 > 1$.
Furthermore, because $y$ and $z$ are symmetry directions, $n_y$ and $n_z$ stay constant thoughout the problem.
Thus, we must have $n_\perp^2 < 0$ in the vacuum, implying that $n_{x}$ is imaginary, and the wave is evanescent.
This switching of the wave from evanescent to propagating at the boundary reflects the crossing of the $P = 0$ cutoff at the boundary \cite{stix1992waves}.

For the evanescent wave in the vacuum, we can decompose the electric field into an ``incident'' wave with $\Im (n_{xI}) > 0$, and a ``reflected'' wave with $n_{xR} = - n_{xI}$.
Thus, the complex amplitudes of the vacuum wave components take the form:
\begin{align}
	\tE_{I} & = \frac{n_{xI}}{n_{\perp I}} E_{0I} \hat{x} + \frac{n_{y}}{n_{\perp I}} E_{0I} \hat{y} - \frac{n_{\perp I}}{P_1 n_z} E_{0I} \hat{z}\\
	\tE_{R} & = -\frac{n_{xR}}{n_{\perp I}} E_{0R} \hat{x} + \frac{n_{y}}{n_{\perp I}} E_{0R} \hat{y} - \frac{n_{\perp I}}{P_1 n_z} E_{0R} \hat{z}.
\end{align}
In the plasma region, there will be a single propagating ``transmitted'' wave, with $n_T > 0$, with complex wave component amplitudes:
\begin{align}
	\tE_{T} & = \frac{n_{xT}}{n_{\perp T}} E_{0T} \hat{x} + \frac{n_{y}}{n_{\perp T}} E_{0T} \hat{y} - \frac{n_{\perp T}}{P_2 n_z} E_{0T} \hat{z}. \label{eq:Ect}
\end{align}
As before, the magnetic field amplitude is given from the electric field amplitude in each case by $\ve{\tB} = \ve{n} \times \ve{\tE}$.

For a boundary in the $y$-$z$ plane at $x=0$, the boundary matching conditions for materials with magnetic permeability $\mu = 1$ are \cite{Griffiths2017,Jackson1999}:
\begin{align}
	\epsilon_{1,xj} \tE_{1}^j &= \epsilon_{2,xj} \tE_{2}^j  \label{eq:boundaryMatchEperp}\\
	\tE_{1}^i &= \tE_{2}^i \qquad \; i \in \{y,z\} \label{eq:boundaryMatchEparallel}\\
	\tB_{1}^i &= \tB_{2}^i \qquad \; i \in \{x,y,z\}. \label{eq:boundaryMatchB}
\end{align}
Plugging our waveforms into these equations, Eqs.~(\ref{eq:boundaryMatchEparallel}) and the $x$ component of Eq.~(\ref{eq:boundaryMatchB}) yield:
\begin{align}
	E_{0I} + E_{0R} &= \alpha E_{0T}; \quad \alpha \equiv \frac{n_{\perp I}}{n_{\perp T}},
\end{align}
while Eq.~(\ref{eq:boundaryMatchEperp}) and the $y$ component of Eq.~(\ref{eq:boundaryMatchB}) yield:
\begin{align}
	E_{0I} - E_{0R} &= \beta E_{0T}; \quad \beta \equiv \frac{n_{xT}}{n_{xI}} \frac{n_{\perp I}}{n_{\perp T}}.
\end{align}
The $z$ component of Eq.~(\ref{eq:boundaryMatchB}) is trivially satisfied, since $\tB^z = 0$.
Note that $\alpha$ is a pure imaginary number, while $\beta$ is a pure real number.
The solution to these coupled equations is:
\begin{align}
	E_{0R} &= \frac{\alpha - \beta}{\alpha + \beta} E_{0I} \label{eq:Er}\\
	E_{0T} &= \frac{2}{\alpha - \beta} E_{0I} \label{eq:Et}.
\end{align}
These equations take the exact same form as that for $p$-polarized light for an isotropic dielectric in introductory electromagnetism \cite{Griffiths2017}, except for the redefinition of $\alpha$ and $\beta$.

\subsection{Electromagnetic energy flux in the vacuum}

We are now in a position to calculate the electromagnetic energy flux in the vacuum. 
Since $\tB^z = 0$, we have:
\begin{align}
	S^x_{EM} &= - \frac{c}{4\pi} \llangle E_z B_y \rrangle.
\end{align}
The relevant fields are given by:
\begin{small}
\begin{align}
	E_z &= \Re \biggl[ -\frac{n_{\perp I}}{n_z} \biggl( E_{0 I} e^{i k_{xI} x} + E_{0 R} e^{-i k_{xI} x} \biggr) e^{i k_y y +i k_z z - i\omega t}   \biggr]\\
	B_y &= \Re \biggl[ \frac{n_{xI}}{n_z n_{\perp I}} \biggl( E_{0 I} e^{i k_{xI} x} - E_{0 R} e^{-i k_{xI} x} \biggr) e^{i k_y y +i k_z z - i\omega t}   \biggr].
\end{align}
\end{small}
The average of two quantities $A$ and $B$ oscillating at the same frequency is given by $\llangle A B \rrangle = \Re(A^* B) /2$.
Thus:
\begin{small}
\begin{align}
	S^x_{EM} &= \frac{c}{8\pi} \Re \biggl\{ \biggl[ \frac{n_{\perp I}}{n_z} \biggl( E_{0 I} e^{i k_{xI} x} + E_{0 R} e^{-i k_{xI} x} \biggr) \biggr]^* \notag\\
	& \hspace{0.25in}\times\biggl[ \frac{n_{xI}}{n_z n_{\perp I}} \biggl( E_{0 I} e^{i k_{xI} x} - E_{0 R} e^{-i k_{xI} x} \biggr) \biggr] \biggr\}\\
	&= \frac{c}{8\pi} \Re \biggl\{ \frac{n_{xI} n_{\perp I}^*}{n_z^2 n_{\perp I}}\notag\\
	& \hspace{0.15in} \times \biggl[ \biggl( E_{0 I} E_{0I}^* e^{-2 \Im(k_x) x} - E_{0 R} E_{0R}^* e^{2 \Im(k_x) x} \biggr)  \notag\\
	& \hspace{0.15in} + \biggl( -E_{0 I}^* E_{0R} e^{-2 i \Re(k_x) x} + E_{0 I} E_{0R}^* e^{2 i \Re(k_x) x} \biggr) \biggr] \biggr\}.
\end{align}
\end{small}
Here, we have separated out the terms in parentheses; the first parentheses contain a purely real quantity, and the second parenthese contain a purely imaginary quantity.
For a propagating wave in vacuum (as is considered in boundary value problems in introductory electromagnetism), $n_{xI}$ is real, and it is the first set of parentheses that matter; we can recognize these as the energy fluxes associated with the incident and reflected waves, respectively.
However, when $n_{xI}$ is imaginary, as for our vacuum evanescent wave, it is the cross amplitudes in the second set of parentheses that determine the energy flux.
Thus, using the fact that $n_{xI}$ and $n_{\perp I}$ are pure imaginary, we have:
\begin{align}
	S^x_{EM} &= -\frac{c}{4\pi n_z^2} \Im (n_{xI})\Im \lp E_{0 I}^* E_{0R} \rp .
\end{align}

Now we make use of Eq.~(\ref{eq:Er}), and the fact that $\alpha$ is imaginary and $\beta$ is real.
We have:
\begin{align}
	\Im \lp E_{0 I}^* E_{0R} \rp &= \Im \lp E_{0 I}^* E_{0I} \frac{\alpha - \beta}{\alpha + \beta}\rp\\
	&= \Im \lp \alpha  \rp \frac{2 \beta}{\beta^2 - \alpha^2} E_{0 I}^* E_{0I}.
\end{align}
Plugging this back in, using the definitions of $\alpha$ and $\beta$ and the fact that $n_{\perp T}^2/n_{\perp I}^2=P_2$, yields:
\begin{align}
	S^x_{EM} &= \frac{c}{2\pi n_z^2} \frac{n_{xT}}{P} \frac{E_{0 I}^* E_{0I}}{\beta^2 -\alpha^2}  .
\end{align}

\subsection{Minkowski energy flux in the plasma}

The Minkowski energy flux in the plasma is given from Eq.~(\ref{eq:sMinkowski}) by:
\begin{align}
	S^x_M &= \omega \mathcal{I} v_g^x.
\end{align}
The group velocity is given by $v_g^x = -(\paf{D}{k_x})/(\paf{D}{\omega})$, with 
\begin{align}
	D &= k_x^2 + k_y^2 + \lp 1 - \frac{\omega_{pe}^2}{\omega^2} \rp \lp n_z^2 - 1 \rp,
\end{align}
which yields:
\begin{align}
	v_g^x = P c n_{xT} \lp P^2 n_z^2 + n_{\perp T}^2 \rp^{-1}.
\end{align}
From Eq.~(\ref{eq:actionElectromagnetic}), we have
\begin{align}
	\mathcal{I} &= \frac{1}{16 \pi \omega^2} \tE_{T}^{i*} \pa{}{\omega_r} \lp \omega^2 \bvec 1 & 0 & 0\\ 0 & 1 & 0 \\ 0 & 0 & 1-\frac{\omega_{pe}^2}{\omega^2}\evec \rp \tE^j\\
	&= \frac{1}{8 \pi \omega} \tE_{T}^{i*} \tE_{T}^j.
\end{align}
Plugging in our definition for $\tE_{T}$ in Eq.~(\ref{eq:Ect}), and recalling that $n_{xT}$ is real, we find:
\begin{align}
	\mathcal{I} &= \frac{1}{8 \pi \omega} \frac{P^2 n_z^2 + n_{\perp T}^2}{n_z^2 P^2} \tE_{0T}^* \tE_{0T}.
\end{align}

We can evaluate $E_{cT}^* E_{cT}$ using Eq.~(\ref{eq:Et}).
Recalling that $\alpha$ is imaginary and $\beta$ real, this gives:
\begin{align}
	E_{0T}^* E_{0T} &= \frac{4}{\beta^2 - \alpha^2} E_{0I}^* E_{0I}.
\end{align}
Putting this all together, we find:
\begin{align}
	S^x_{M} &= \frac{c}{2\pi n_z^2} \frac{n_{xT}}{P} \frac{E_{0 I}^* E_{0I}}{\beta^2 -\alpha^2} ,
\end{align}
so that $S^x_{M} = S^x_{EM}$.

\subsection{Energy-momentum flow to resonant particles}

From Eq.~(\ref{eq:actionConservation}), we have:
\begin{align}
	\pa{}{x} \lp v_g^i \mathcal{I} \rp= -\Gamma \mathcal{I}.
\end{align}
Thus, as the wave travels from the dilute plasma region to the mode conversion layer, the quantity $v_g^i \mathcal{I}$ remains constant.
At the mode conversion layer, the action flux magnitude is conserved, but the direction flips sign.
Then, $v_g^i \mathcal{I}$ is constant again until the edge of the uniform region with the resonant particles.
Because $\omega$, $k_y$, and $k_z$ are constant throughout, this means in turn that the final quantities (upon entering the uniform region) $S^x_{Mf}$, $\Pi^{yx}_{Mf}$, and $\Pi^{zx}_{Mf}$ are equal in magnitude and opposite to their initial signs at the interface with the vacuum.

In the uniform region with resonant particles, the action starts to spatially decay, and eventually disappears.
However, we know that over this region, the conservation Eq.~(\ref{eq:consMRP}) holds.
We can calculate the energy transfer rate to resonant particles by taking the flux terms $\ve{S}_{RP}$ and $\ve{\Pi}_{RP}$ in the resonant particle EMT to be 0.
Then, Eq.~(\ref{eq:consMRP}) becomes:
\begin{align}
	\pa{}{t} \bvec W_{RP} \\ p_{RP}^y \\ p_{RP}^z \evec &= -\pa{}{x} \bvec S_{Mf}^x \\ \Pi_{Mf}^{yx} \\ \Pi_{Mf}^{zx} \evec.
\end{align}
Now, we integrate over the uniform region volume, assuming that the damping is strong enough that the wave has completely damped out by the low-$x$ edge of the uniform region.
Thus, integrating over an area $A$ in the $y$-$z$ plane, and using our derived relations between the Minkowski and electromagnetic momentum, we fine:
\begin{align}
	\pa{}{t} \bvec U_{RP} \\ P_{RP}^y \\ P_{RP}^z \evec &= A \bvec S_{EM}^x \\ \Pi_{EM}^{yx} \\ \Pi_{EM}^{zx} \evec,
\end{align}
where $U_{RP}$ and $\ve{P}_{RP}$ are the volume-integrated resonant particle energies and momenta respectively.
Thus, we see that in the boundary-value problem, the energy and momentum that end up in the resonant particles are ultimately supplied by the electromagnetic field.

Now, it is still possible that in spite of this, there is a response of nonresonant particles in the plasma to the wave.
However, because the electromagnetic momentum and energy that enters the plasma is all accounted for in the resonant particles, such a response could only lead to a \emph{rearrangement} of energy and momentum within the wave region.
Thus, the \emph{net force} (or net torque in a cylindrically symmetric system) on the plasma all results from flow of electromagnetic momentum through the vacuum bordering the plasma, and is consistent with the total force / torque on the resonant particles.

While it is important to know the net force on the plasma volume, it is also often important to know the local force, and thus to see if a momentum rearrangement within the plasma due to a nonresonant response does in fact take place.
In the next section, we will show that no such momentum rearrangement takes place, at least within the uniform region.
Thus, the local force on the resonant particles will be shown to constitute the total local force on the plasma.

\section{Warm-fluid model of the electrostatic wave} \label{sec:WarmFluidModel}

We will now shift our focus to the uniform regime in Fig.~(\ref{fig:waveInjectionModel}), and the mode-converted electrostatic wave that damps on the resonant particles there.
We will employ a warm-fluid model to describe the bulk plasma response.
Of course, the resonant particles cannot be described by this fluid model; thus, we will assume that the resonant particle damping is calculated already, and appears as a ``given'' imaginary portion of the dispersion relation.
In light of Eq.~(\ref{eq:consMRP}), this information immediately tells us the energy and momentum transfer to the resonant particles.
Our focus in this section will be the momentum transferred to the \emph{nonresonant} particles at the same time.

Our goal is to show that the momentum conservation principle from the closed system in Eq.~(\ref{eq:consEMP}) is ultimately consistent with the absence of a nonresonant force along the symmetry directions in steady-state lower hybrid current and rotation drive.
Crucially, we want to accomplish this in a theory that also captures the momentum cancellation result for the initial value problem.
We will thus proceed fairly slowly, calculating each component of the tensor from the first principles presented, and showing that the familiar fluid equations respect this momentum conservation.
We will then perform a quasilinear analysis of these equations to show how the nonresonant force vanishes, leaving only the resonant force.

Our analysis here is similar in spirit to that in Refs.~\cite{Diamond1991} and \cite{Gao2006,Gao2007}.
However, in contrast to the former, our analysis will incorporate wave dissipation and the transfer of energy and momentum to the resonant particles, and will not assume an $\ve{E}\times \ve{B}$ radial fluid velocity.
In contrast to the latter, our warm fluid analysis gives a finite group velocity, which allows us to relate the \emph{spatial action gradients} to the magnitude of the damping.
This in turn enables a comparison of the resonant forces, which depend on the damping, to the nonresonant forces, which depend on the gradients.
The vanishing of the nonresonant response in the boundary-value problem thus requires evaluating the problem to this order.

\subsection{Preliminaries: electrostatic wave dispersion}

For any electrostatic wave, our starting point is the Poisson equation:
\begin{align}
	-\nabla^2 \phi = 4\pi \sum_s q_s n_s.
\end{align}
Generally, we assume quasineutrality, wherein the 0th-order charge densities of the various species cancel.
Thus, the charge density is determined by the first-order densities $n_1$:
\begin{align}
	-\nabla^2 \phi_1 = 4\pi \sum_s q_s n_{s1}. \label{eq:poissonLinearized}
\end{align}
Fourier transforming in $x$ and $t$, we find
\begin{align}
	k^2 \tphi = 4\pi \sum_s q_s \tn_{s}.
\end{align}
This gives us the dispersion function:
\begin{align}
	D &\equiv 1 + \sum_s D_s \\
	D_s &\equiv -\frac{4 \pi q_s }{k^2} \frac{\tn_s}{\tphi}. \label{eq:ESDispersionGeneral}
\end{align}
To get $D_s$, we will in general have to solve the fluid or Vlasov equations; however, for deriving the general form of the force on the plasma in terms of the dispersion relation, the above form is sufficient.

It will be useful to separate the dispersion function into components that are real ($D_r$)  and imaginary ($D_i$) when evaluated at real $\omega$ and $k$.
We define $\omega_i \equiv \Im(\omega)$, $\boldsymbol{\kappa} \equiv \Im(k)$. 
In the eikonal limit $|\omega_i / \omega_r| \ll 1$, $|\kappa_i / k_i| \ll 1$, our dispersion relation becomes:
\begin{align}
	0 &= D_r \\
	& = 1 + \sum_s D_{rs} \label{eq:Dr2d}\\
	0 &=  \lp \omega_i \pa{}{\omega_r} + \boldsymbol{\kappa} \cdot \pa{}{\ve{k}} \rp D_{r} + D_{i} \label{eq:Di2d1}\\
	&= \sum_s \left[\lp \omega_i \pa{}{\omega_r} + \boldsymbol{\kappa} \cdot \pa{}{\ve{k}} \rp D_{rs} + D_{is}\right], \label{eq:Di2d}
\end{align}
where $D_{rs}$ and $D_{is}$ are the real and imaginary parts of $D_s$ evaluated at real $\omega$ and $k$.
Because $\mathcal{I} \sim |\phi|^2$, we have $\paf{\mathcal{I}}{t} = 2\omega_i \mathcal{I}$ and $\paf{\mathcal{I}}{x^i} = -2\kappa^i \mathcal{I}$, and Eq.~(102) can be seen to be equivalent to Eqs.~(\ref{eq:actionConservation}-\ref{eq:actionDissipationGamma}).

We note that since $k$ and $\omega$ now have imaginary components, there can be some confusion in the definition of the tilde quantites from Eq.~(\ref{eq:EfromEc}).
We will choose the convention that includes the imaginary parts:
\begin{align}
	\phi = \Re \lp \ve{\tphi} e^{i\ve{k}_r \cdot \ve{x} - i\omega_r t} e^{-\ve{\kappa} \cdot \ve{x} + \omega_i t} \rp.
\end{align}
Thus, the local amplitude $\phi_a$ of the wave will be given by:
\begin{align}
	|\phi_a| &= |\ve{\tphi}| e^{-\ve{\kappa} \cdot \ve{x} + \omega_i t},
\end{align}
where $\tphi$ is constant in space and time.
Then, the electromagnetic energy density is given by
\begin{align}
	W_{EM} = \frac{k_r^2 |\phi_a|^2}{16\pi} \propto e^{-2\ve{\kappa} \cdot \ve{x} + 2\omega_i t}. \label{eq:wEMElectrostaticWave}
\end{align}
This convention makes taking derivatives straightforward, but can be a little confusing.

\subsection{Wave action and resonant particle force}

We will start by calculating the wave action and resonant particle force in the electrostatic theory.
This will allow us to clearly disambiguate resonant and nonresonant forces later in the problem.

The wave action is given from Eq.~(\ref{eq:actionElectromagnetic}) by:
\begin{align}
	\mathcal{I} &= \frac{1}{16 \pi \omega^2} \tE^{i*} \pa{}{\omega_r} \lp \omega^2 \epsilon_{H,ij} \rp \tE^j e^{-2\ve{\kappa} \cdot \ve{x}+2\omega_i t}. 
\end{align}
To calculate this for electrostatic waves, we will have to relate the species susceptibility $\chi_H^{ij}$ to the dispersion function $D_{rs}$.
Using the Fourier-transformed charge continuity equation, and the definition \cite{stix1992waves} of the susceptibility $\tilde{j}_i = -i \omega \chi_{H,ij} \tE^j / 4 \pi $, we have:
\begin{align}
	k^2 D_{rs} \tphi &= 4 \pi \tilde{\rho}_s = -4\pi\frac{k^m \tilde{j}_m}{\omega} = k^m \chi_{H,mn} k^n \tphi.
\end{align}
Using this in to our action equation, we find:
\begin{align}
	\mathcal{I} &= W_{EM} \sum_s \pa{D_{rs}}{\omega} = W_{EM} \pa{D_r}{\omega}. \label{eq:actionElectrostatic}
\end{align}

Consider the conserved Minkowski-resonant particle system; specifically, the $i \in (y,z)$ components of Eq.~(\ref{eq:stressTensorDiv0}), for $\ve{T}$ in Eq.~(\ref{eq:consMRP}).
Using Eqs.~(\ref{eq:groupVelocity}), (\ref{eq:actionElectrostatic}), and (\ref{eq:Di2d1}), we find the simple result:
\begin{align}
	\pa{p_{RP}^i}{t} &= - \pa{p_M}{t} - \pa{}{x^j} \Pi^{ij}_M \\
	&= -k^i \lp \pa{\mathcal{I}}{t} + \pa{}{x^j} \lp v_g^j \mathcal{I} \rp \rp \\
	&= -2 k^i \lp \omega_i + \kappa_j \frac{\paf{D_{r}}{k^j}}{\paf{D_{r}}{\omega}}  \rp \lp W_{EM} \pa{D_r}{\omega} \rp\\
	&= 2 W_{EM} k^i D_i.
\end{align} 
It is then clear that the species-specific resonant force is:
\begin{align}
	\pa{p_{RP,s}^i}{t} &= 2 W_{EM} k^i D_{is}.
\end{align}

\subsection{EMT-consistent force from oscillating electric field}

With the force on the resonant particles established, we now turn to the momentum conservation-consistent total force on the particle distribution.
The total electromagnetic force on the plasma is given from the closed system Eq.~(\ref{eq:consEMP}) as:
\begin{align}
	\pa{p_{P}^i}{t} + \pa{}{x^j} \Pi_P^{ij} &= -\pa{p_{EM}^i}{t} - \pa{}{x^j} \Pi_{EM}^{ij}.
\end{align}
Now, an electrostatic wave does technically have a small magnetic field associated with it, and thus nonvanishing momentum $p_{EM}^i$.
This can be seen by Lorentz boosting the truly electrostatic solution in the frame traveling at the wave phase velocity to the observer frame.
However, for a wave traveling at phase velocity $v_p = |\omega_r/k_r| \ll c$, this magnetic field is $\mathcal{O}(v_p/c)$ smaller than the electric field, and thus the momentum from Eq.~(\ref{eq:pEM}) is $\mathcal{O}(v_p^2/c^2)$ smaller than the stress terms, and can be ignored.
Thus, the total force from the wave on the plasma, which we denote $\ve{F}_{EM}$, can be written:
\begin{align}
	F_{EM}^i &= - \pa{}{x^j} \Pi_{EM}^{ij}.
\end{align}

We are ultimately interested in the average effect of the wave on the plasma, rather than the oscillations themselves.
This requires calculating the average value of the electromagnetic stress tensor over one of the symmetry directions $y$ or $z$.
To lowest order in $|\kappa / k|$, we have:
\begin{align}
	\llangle \Pi_{EM}^{ij} \rrangle_y &= -\frac{1}{4\pi} \llangle E^i E^j - \frac{1}{2} \delta^{ij} E^2 \rrangle_y\\
	&= -\frac{e^{-2\ve{\kappa} \cdot \ve{x} + 2\omega_i t}}{8\pi} \Re \left[ \tE^{i*} \tE^{j} - \frac{1}{2} \delta^{ij} \tE^{m*} \tE_{m} \right]\\
	&= -2 \frac{W_{EM}}{k_r^2} \lp k_r^i k_r^j - \frac{1}{2} \delta^{ij} k_r^2 \rp,
\end{align}
where we used $\tE^i = - i k^i \tphi$ and Eq.~(\ref{eq:wEMElectrostaticWave}).
Taking the derivative to obtain the force thus yields (using the scaling in Eq.~(\ref{eq:wEMElectrostaticWave})):
\begin{align}
	\llangle F_{EM}^i \rrangle &= -4 \frac{W_{EM}}{k_r^2} \kappa_j \lp k_r^i k_r^j - \frac{1}{2} \delta^{ij} k_r^2 \rp.
\end{align}

Of course, we are often interested in calculating the specific force $\ve{F}_{EM,s}$ the plasma exerts on each species $s$ in the plasma.
To do this, we need to calculate the average correlation between the density of $s$ and the electric field:
\begin{align}
	\llangle F_{EM,s}^i \rrangle &= \llangle E^i q_s n_s\rrangle\\
	&= \frac{1}{2} \Re \left[\tE^{i*} q_s \tn_s \right]\\
	&= 2 \frac{W_{EM}}{k_r^2} \Im \left[ k^* k^2 D_s \right],
\end{align}
where we have used Eq.~(\ref{eq:ESDispersionGeneral}) and Eq.~(\ref{eq:wEMElectrostaticWave}).
Now, keeping only to lowest order in $\kappa$, we can Taylor expand around real $\omega,k$ as in Eq.~(\ref{eq:Di2d}) to find:
\begin{align}
	\llangle F_{EM,s}^i \rrangle &\approx  2 W_{EM} k_{r}^i \left[ D_{is} + \lp \kappa^\lambda \pa{}{k^\lambda} + \omega_i \pa{}{\omega_r} \rp D_{rs} \right] \notag \\
	& \hspace{0.3in}+ 4 \frac{W}{k_r^2} \kappa_j \lp k_r^i k_r^j - \frac{1}{2} k_r^2 \delta^{ij} \rp D_{rs}. \label{eq:FEms}
\end{align}
Here, the term with $D_{is}$ is the resonant force, and all other terms are the nonresonant force.
If we sum this over all species and make use of Eqs.~(\ref{eq:Dr2d}) and (\ref{eq:Di2d}), we find $\llangle \ve{F}_{EM} \rrangle = \sum_s \llangle \ve{F}_{EM,s} \rrangle$, showing that this electric force is (as expected) consistent with the momentum conservation law.

Eq.~(\ref{eq:FEms}) also shows why a warm fluid model is, in general, necessary to demonstrate that vanishing of the nonresonant ponderomotive force to the order of the resonant force.
Consider a wave interacting with only a single species.
In steady state, from Eq.~(\ref{eq:Di2d}), we see that for this wave:
\begin{align}
	D_{is} &= -\kappa^\lambda \pa{D_{rs}}{k^\lambda}.
\end{align}
Thus, the nonresonant force contribution from the  of the same order nonresonant force term involving $\paf{D_{rs}}{k^j}$, which is usually 0 in the cold fluid model.
Thus the warm fluid model is actually the simplest model which can calculate the nonresonant force to the correct order.

\subsection{Fluid equations from particle EMT}

In order to calculate the total force on a plasma fluid element, we plug a fluid ansatz into the EMT components for the particle distribution in Eqs.~(\ref{eq:wP}-\ref{eq:piP}).
The ansatz we use is typically a position-dependent Maxwellian:
\begin{align}
	f_s(\ve{x},\ve{v}) &= \frac{n(\ve{x})}{(2 \pi T_s(\ve{x})/m_s)^{3/2}} e^{m_s (\vv-\ve{u}_s(\ve{x}))^2 / 2 T_s(x)},
\end{align}
though this shape is not essential so much as the fact that the distribution is primarily located in a region with $v \ll v_p$.

Plugging this ansatz into Eqs.~(\ref{eq:wP}-\ref{eq:sP}), we find to lowest order in $m c^2$ that for species $s$:
\begin{align}
	W_{Ps} &= m_s c^2 n_s\\
	S_{Ps} &= m_s c^2 n_s u_s^i.
\end{align}
The EMT conservation Eq.~(\ref{eq:stressTensorDiv0}) for component $i=0$ thus gives a mass-weighted sum over the familiar fluid continuity equations for each species:
\begin{align}
	\sum_s m_s \left[\pa{n_s}{t}  + \pa{}{x^j} (n_s u_s^j) \right]&= 0. \label{eq:fluidContinuity}
\end{align}
Of course, we take these each to be individually satisfied.

We also have:
\begin{align}
	p_{Ps} &= m_s n_s u_s^i\\
	\Pi^{ij}_{Ps} &= m_s n_s u_s^i u_s^j + P_s \delta^{ij}, 
\end{align}
with $P_s = n_s T_s$, so that the EMT conservation Eq.~(\ref{eq:stressTensorDiv0}) for components $i=1$-$3$ gives the sum over the momentum equations:
\begin{align}
	\sum_s \left[\pa{}{t}(m_s n_s u_s^i) + \pa{}{x^j} (n_s u_s^i u_s^j) + \pa{P_s}{x^i} \right]&= -\pa{}{x^j} \Pi^{ij}_{EM}.
\end{align}
Because the separate species only interact through the influence of the electric field, and because the electric forces sum to the electromagnetic stress tensor, this equation is simply the sum of the individual momentum equations:
\begin{align}
	\pa{}{t}(m_s n_s u^i) + \pa{}{x^j} (n_s u^i u^j) + \pa{P_s}{x_i} &= F_{EM,s}^i . \label{eq:fluidMomentum1}
\end{align}
Often, this is combined with the continuity equation to obtain:
\begin{align}
	m_s n_s \lp \pa{}{t} + u_s^j \pa{}{x^j} \rp u_s^i &= F_{EM,s}^i \label{eq:fluidMomentum2}.
\end{align}
Thus, the standard fluid momentum equation is consistent with our momentum conservation law, as expressed in the closed system in Eq.~(\ref{eq:consEMP}).

We close this system of equations with an adiabatic expression for the pressure evolution:
\begin{align}
	\lp \pa{}{t} + u^j \pa{}{x^j} \rp \lp \frac{P_s}{n_s^{\gamma_s}} \rp &= 0, \label{eq:fluidClosure}
\end{align}
where $\gamma_s$ is the adiabatic index for species $s$.

\subsection{Linearizing the fluid equations}

To study waves, we must linearize these equations.
To clean the notation, we will suppress the subscripts $s$ in this section.
Now when we linearize, we take the standard approach of decomposing into an average contribution $n_0$, $\ve{u}_0$, and $P_0$, and a smaller oscillating portion $n_1$, $\ve{u}_1$, and $P_1$.
[Interestingly, you get slightly different results if you use, e.g., $n_1$, $\ve{p}_1$, and $T_1$].
For simplicity, we take $u_0 = 0$, corresponding to the application of torque to a non-rotating plasma; however, the results can be easily generalized, if desired.

To first order, from Eqs.~(\ref{eq:fluidContinuity}), (\ref{eq:fluidMomentum2}), and (\ref{eq:fluidContinuity}), we get the familiar warm fluid equations \cite{stix1992waves}:
\begin{align}
	\pa{n_1}{t} &= - n_0 \pa{u_1^j}{x^j} \label{eq:fluidContinuityRest}\\
	m n_0 \pa{u_1^i}{t}  &=  -\pa{P_1}{x^i} + q n \lp E^i +\frac{1}{c}\epsilon^{ijk} u_{1j} B_{0k} \rp \label{eq:fluidMomentumRest}\\
	\pa{P_1}{t} &= \gamma T_0 \pa{n_0}{t},\label{eq:fluidPressureRest}
\end{align}
where $T_0 = \gamma n_0$.

At second order, we obtain our quasilinear theory.
We will use the original form (Eq.~(\ref{eq:fluidMomentum1})) of the momentum equation here, obtaining the average force on the plasma, which we define as the time change in the average momentum:
\begin{align}
	F_\text{fluid} &\equiv \pa{\llangle p_P^i \rrangle}{t}  = -\pa{}{x^j} \Pi^{ij}_\text{Rey} + \llangle F_{EM,s} \rrangle.	
\end{align}
Here, the average momentum in the plasma is given by:
\begin{align}
	\llangle p_P^i \rrangle &=  m n_0 u_0^i + m \llangle n_1 u_1^i \rrangle,
\end{align}
and we have identified the Reynolds stress:
\begin{align}
	\Pi^{ij}_\text{Rey} &= m n_0 \llangle u_1^i u_1^j \rrangle = \frac{1}{2} m n_0 \Re \left[ \tu_1^i \tu_1^j\right] e^{-2\ve{\kappa} \cdot \ve{x} + 2 \omega_i t}\label{eq:ReynoldsStress}
\end{align}
We will show that for the lower hybrid wave in steady state, this Reynolds stress cancels with the electromagnetic force $F_{EM,s}$ in just such a way as to leave only the force on the resonant particles.
Note crucially that the factor of $\paf{}{x^i}$ out front of the stress term implies a factor of the decay rate $\kappa^i$; thus, in order to calculate the force to the same order as the electromagnetic EMT, i.e. first order in $\kappa^i$, we need only calculate the tensor in brackets to 0th order in $|\kappa| / |k|$.

\subsection{Lower hybrid wave solution}

Eqs.~(\ref{eq:poissonLinearized}) and (\ref{eq:fluidContinuityRest}-\ref{eq:fluidPressureRest}) can be Fourier transformed and solved to yield both the dispersion function $D_s$ (from Eq.~(\ref{eq:ESDispersionGeneral})):
\begin{align}
	D_s &= -\frac{\omega_{ps}^2}{k^2} \lp \frac{k_\perp^2}{\omega^2 - \Omega_s^2} +  \frac{k_z^2}{\omega^2} \rp C_s, \label{eq:DsElectrostaticMagnetized}
\end{align}
and fluid velocities $\tu^i_s$ for species $s$:
\begin{align}
	\tu_s^x &= \frac{q_s}{m_s} \frac{k_x \omega + i k_y \Omega_s}{\omega^2 - \Omega_s^2} C_s \label{eq:uxEsMagnetized}\\
	\tu_s^y &= \frac{q_s}{m_s} \frac{k_y \omega - i k_x \Omega_s}{\omega^2 - \Omega_s^2} C_s\\
	\tu_s^z &= \frac{q_s}{m_s} \frac{k_z}{\omega} C_s \label{eq:uzEsMagnetized},
\end{align}
where $k_\perp^2 = k_x^2 + k_y^2$, and $C_s$ captures the thermal corrections and is given by:
\begin{align}
	C_s &= \lp 1 - \frac{\gamma_s (k_x^2 + k_y^2)  v_{ths}^2}{\omega^2-\Omega_z^2} - \frac{\gamma_s  k_z^2 v_{ths}^2}{\omega^2} \rp^{-1},
\end{align}
where $v_{ths} \equiv \sqrt{T_s/m_s}$.

The warm lower hybrid dispersion relation can be recovered if we take $|\Omega_i| \ll |\omega| \ll |\Omega_e|$, $v_{thi} \ll |\omega/k|$, $v_{the} \ll |\Omega_e/k|$, and $|k_z / k| \ll |\omega^2/(\omega^2-\Omega_s^2)| \; \forall s$.
This can be shown to agree with the kinetic dispersion relation when $\gamma_e = 3/4$ and $\gamma_i = 3$ \cite{Verdon2009}.
In this limit,
\begin{align}
	C_s &\approx 1 + \frac{\gamma_s (k_x^2 + k_y^2)  v_{ths}^2}{\omega^2-\Omega_z^2} + \frac{\gamma_s  k_z^2 v_{ths}^2}{\omega^2}.
\end{align}

\subsection{Total force on fluid for magnetized electrostatic wave}

Now we are in a position to calculate the force on a fluid element.
To calculate the quasilinear force, first note that Eq.~(\ref{eq:FEms}) can be rewritten:
\begin{align}
	\llangle F_{EM,s}^i \rrangle &\approx  2 W_{EM} \biggl[ k_{r}^i D_{is} - \kappa^i D_{rs} \notag\\
	&\hspace{0.15in}+  \frac{k^{ir}}{k^2} \lp \kappa^\lambda \pa{}{k^\lambda} + \omega_i \pa{}{\omega_r} \rp \lp k^2 D_{rs} \rp   \biggr].
\end{align}
Focus on the term involving the derivative with respect to $k$:
\begin{align}
	\llangle F_{EM,s,k}^i \rrangle \equiv \frac{k^{ir}}{k^2} \kappa^\lambda \pa{}{k^\lambda} \lp k^2 D_{rs} \rp .
\end{align}
For our problem, we only need to consider $\lambda = x$.
Thus, using Eq.~(\ref{eq:DsElectrostaticMagnetized}), we have:
\begin{align}
	\llangle F_{EM,s,k}^i \rrangle &= -4 W_{EM} \frac{k_{r}^i k_r^x}{k^2} \kappa_x  \frac{\omega_{ps}^2}{\omega^2 - \Omega_s^2} \notag \\
	&\quad \times \lp 1 + 2 \gamma v_{ths}^2 \lp \frac{k_\perp^2}{\omega^2 - \Omega_s^2} + \frac{k_z^2}{\omega^2} \rp \rp .
\end{align}

Now we calculate the Reynold's stress.
Plugging Eqs.~(\ref{eq:uxEsMagnetized}-\ref{eq:uzEsMagnetized}) into Eq.~(\ref{eq:ReynoldsStress}), we find:
\begin{align}
	\Pi_\text{Rey}^{ix} &= 2 W_{EM} \frac{\omega_{ps}^2}{\omega^2 - \Omega_s^2} \left[ \frac{k_i k_x}{k^2} C_s^2 + \delta^{ix} \frac{\Omega_s^2}{(\omega^2 - \Omega_s^2)} \frac{k_\perp^2}{k^2} C_s^2 \right].
\end{align}
Examining the force along the symmetry directions, we take $i \in (y,z)$.
By taking the $x$ derivative using Eq.~(\ref{eq:wEMElectrostaticWave}), we find:
\begin{align}
	-\pa{}{x} \Pi^{ix}_\text{Rey} &= -\llangle F_{EM,s,k}^i \rrangle.
\end{align}
Thus, the total force on the fluid element along the symmetry directions is:
\begin{align}
	\llangle F_{\text{tot},s}^i \rrangle &=  2 W_{EM} k_{r}^i \biggl[ D_{is} -  \omega_i \pa{D_{rs}}{\omega_r}     \biggr]; \hspace{.1in} i \in (y,z). \label{eq:Ftot}
\end{align}

Eq.~(\ref{eq:Ftot}) captures the behavior of the plasma in both the 1D initial value problem, and in the steady-state multidimensional boundary problem.
For the 1D IVP, $\kappa_i = 0$, and from Eq.~(\ref{eq:Di2d1}) we see that the sum of all resonant and nonresonant forces on the plasma is 0.
Thus, the nonresonant particles recoil, canceling out the momentum transferred by the wave to the resonant particles.
For the steady-state multidimensional BVP, meanwhile, $\omega_i =0$, and thus the force on the plasma is precisely equal to the force on the resonant particles; in other words, the recoil response on the nonresonant particles vanishes.
Thus, the behaviors of both the IVP and BVP are shown to arise from a consistent, coherent, energy- and momentum-conserving framework.

\section{Discussion}

The topic of flow drive by waves in plasma, and more broadly of ponderomotive forces in plasma, has been a subject of research for many years.
Thus, we begin our discussion with a comparison of our results to some of the existing literature.

It has long been known that the electromagnetic EMT of a propagating electromagnetic wave in vacuum are the same as the Minkowski momentum of that wave \cite{dodin2012axiomatic}.
Thus, for plasma waves which arise from propagating vacuum electromagnetic waves, such as high-frequency $|\omega| > |\omega_{pe}|$ modes, our result in Sec.~\ref{sec:FresnelModel} would be trivial.
Our result differs from this historical result precisely because the vacuum wave is evanescent rather than eikonal, and thus has no defined Minkowski EMT, so that a comparison of EMTs across regions was required.

Sec.~\ref{sec:WarmFluidModel} bears more similarity to the existing literature.
The key role played by the off-diagonal component of the Reynolds stress has been noted in several papers employing a fluid theory.
In the study of low-frequency electrostatic turbulence, similar results for poloidal flow generation \cite{Diamond1991} and parallel momentum transport \cite{Diamond2008} have been obtained; however, an early theorem used in these results relied on assuming an $\ve{E} \times \ve{B}$ radial velocity $u_s^r = E_\theta B_z$, which is certainly not the case for the ions in the lower-hybrid waves we considered here.
In addition, consideration of the different forces on resonant and nonresonant effects are only considered in the paper on parallel momentum transport \cite{Diamond2008}, not the paper on poloidal momentum damping \cite{Diamond1991}, making it difficult to compare the results to theories incorporating only the resonant particles.
Here, we have clearly distinguished resonant and nonresonant forces for both the parallel and perpendicular forces, making clear the deep parallels between current drive and rotation drive via cross-field charge extraction.

The importance of the Reynolds stress was also noted in several papers examining nonresonant current and flow drive in the cold-fluid theory \cite{Gao2006,Gao2007}.
These papers established the cancellation between the electromagnetic force on the nonresonant particles and the Reynolds' stress to zeroth order in $v_{th}^2$.
However, as discussed in Sec.~\ref{sec:WarmFluidModel}, the resonant forces actually appear at $\mathcal{O}(v_{th}^2)$, so it is necessary to work to this order to establish that the nonresonant forces vanish in steady state.
In addition, these references did not consider the possibility of time-dependence in the problem, and so could not show consistency with the initial value problem result.

In addition to fluid theory, the Reynolds stress has appeared in magnetized hot-plasma kinetic theories as well \cite{Berry1999,Jaeger2000,Myra2000,Myra2002,Myra2004}.
These theories do not require a guess for the adiabatic index $\gamma_s$, and are capable of tackling waves, such as Bernstein waves, which do not satisfy the ordering requirements for fluid waves.
However, this ability comes at the price of significant computational complexity, since calculating $\ve{\Pi}_\text{Rey}$ requires evaluating the evolution of the second-order distribution function $f_2$ in a multi-dimensional magnetized plasma.
Thus, none of these papers attempt the time-dependent initial value problem, and thus cannot establish consistency with the conventional quasilinear theory of Landau damping.
Even with this simplification, the calculations are plagued with subtle difficulties.
For instance, it was only recently established \cite{Lee2012,guan2013plasma,guan2013toroidal} that the classic Kennel-Engelman theory of quasilinear diffusion incorrectly gyro-averaged the diffusion tensor rather than the whole quasilinear equation, and thus missed perpendicular momentum input into the resonant particles.
Similarly, the sudden turning on of the wave fields at $t=0$ used to calculate $f_2$ in Refs.~\cite{Berry1999,Jaeger2000} was shown to yield incorrect forces in the cold-fluid limit, an error ultimately coming from the fact that $|\omega_i / \omega_r| \ll 1$ was not satisfied \cite{Chen2013}.

In summation, this paper fulfills a valuable role, providing a relatively simple theory that captures the behaviors of both the initial- and boundary-value problems in magnetized and unmagnetized plasmas, while establishing that when nonresonant forces vanish, they do so at the order of the resonant forces.

In identifying this gap in the literature, it is important to note that we do not claim that existing theories of hot particle extraction by alpha channeling \cite{Fisch1992,fisch1992current,Heikkinen1996,ochs2015coupling,ochs2015alpha,Valeo1994,fisch1995ibw,Fisch1995a,Heikkinen1995,Marchenko1998,Kuley2011,Sasaki2011,Chen2016a,Gorelenkov2016,Cook2017,Castaldo2019,Romanelli2020,Herrmann1997,Cianfrani2018,Cianfrani2019,White2021} are incorrect.
These theories simply focused exclusively on the resonant particles, and thus were not positioned to answer questions that depended sensitively on the response of the nonresonant particles.
For those papers which \emph{did} focus on rotation drive  while neglecting nonresonant particles \cite{fetterman2008alpha,fetterman2012wave}, it turns out fortuitously that the relevant nonresonant response vanishes in steady-state, making the alpha channeling rotation drive scheme possible in practice.

Finally, we note that there is a whole other approach to the calculation of nonresonant ponderomotive forces, using the variational theory of the oscillation center \cite{Dewar1973}.
Such methods begin with an action principle for the single particle, and then transform to a set of oscillation-center coordinates.
These methods are deep and powerful, especially when combined with Weyl transform methods that allow for consistent handling of plasma nonuniformities \cite{Tracy2014}, and come with self-consistent energy and momentum conservation theorems built in.
However, such theories are intrinsically kinetic, and self-consistency involves a Lagrangian coordinate calculation of the oscillation center dynamics.
In addition, the physical currents present in the system are often buried in the transformation from physical to oscillation center coordinates, and can be difficult to extract and integrate over the plasma distribution.
A detailed comparison of cross-field charge transport in the oscillation-center and warm fluid pictures is outside the scope of the present paper, and will be left for future work.

\section{Conclusion}

In this paper, we have shown that the conventional explanation for momentum conservation in steady-state current drive applies only in the case of an initial value problem, which is generally not the case of interest in present devices, in which wave power is brought into the device from a boundary.
In steady state, the nonresonant particle recoil response does not get transferred into ions via collisions, or even get transferred directly into the ions by the wave; it simply does not exist.
This absence of a nonresonant response was shown to be consistent with global energy-momentum conservation, through the use of a Fresnel model which identified the electromagnetic energy and momentum flux in the vacuum with the Minkowski momentum in the plasma.
It was also shown to be consistent with local energy-momentum conservation, via the use of a simple electrostatic warm fluid model.
This model recovered both the behavior of the 1D initial value problem, where a recoil reaction does exist, and the multidimensional steady-state boundary-value problem, where no recoil occurs.
The absence of the recoil response in the boundary value problem allows not only for current drive, but also for the extraction of the charge associated with the resonant particles, and thus for rotation drive via alpha channeling, along with all the advantages rotating plasmas provide.

\section{Acknowledgments:}
We would like to thank E.J. Kolmes and M.E. Mlodik for helpful discussions.
This work was supported by grants NNSA DE-NA0003871, NNSA DE-SC0021248, and DE-SC0016072.



\input{bvp_longPaper.bbl}

\clearpage

\end{document}

%% file: bvp_longPaper.bbl
%

%% file: bvp_longPaper.bbl
\begin{thebibliography}{82}%
\makeatletter
\providecommand \@ifxundefined [1]{%
 \@ifx{#1\undefined}
}%
\providecommand \@ifnum [1]{%
 \ifnum #1\expandafter \@firstoftwo
 \else \expandafter \@secondoftwo
 \fi
}%
\providecommand \@ifx [1]{%
 \ifx #1\expandafter \@firstoftwo
 \else \expandafter \@secondoftwo
 \fi
}%
\providecommand \natexlab [1]{#1}%
\providecommand \enquote  [1]{``#1''}%
\providecommand \bibnamefont  [1]{#1}%
\providecommand \bibfnamefont [1]{#1}%
\providecommand \citenamefont [1]{#1}%
\providecommand \href@noop [0]{\@secondoftwo}%
\providecommand \href [0]{\begingroup \@sanitize@url \@href}%
\providecommand \@href[1]{\@@startlink{#1}\@@href}%
\providecommand \@@href[1]{\endgroup#1\@@endlink}%
\providecommand \@sanitize@url [0]{\catcode `\\12\catcode `\$12\catcode
  `\&12\catcode `\#12\catcode `\^12\catcode `\_12\catcode `\%12\relax}%
\providecommand \@@startlink[1]{}%
\providecommand \@@endlink[0]{}%
\providecommand \url  [0]{\begingroup\@sanitize@url \@url }%
\providecommand \@url [1]{\endgroup\@href {#1}{\urlprefix }}%
\providecommand \urlprefix  [0]{URL }%
\providecommand \Eprint [0]{\href }%
\providecommand \doibase [0]{http://dx.doi.org/}%
\providecommand \selectlanguage [0]{\@gobble}%
\providecommand \bibinfo  [0]{\@secondoftwo}%
\providecommand \bibfield  [0]{\@secondoftwo}%
\providecommand \translation [1]{[#1]}%
\providecommand \BibitemOpen [0]{}%
\providecommand \bibitemStop [0]{}%
\providecommand \bibitemNoStop [0]{.\EOS\space}%
\providecommand \EOS [0]{\spacefactor3000\relax}%
\providecommand \BibitemShut  [1]{\csname bibitem#1\endcsname}%
\let\auto@bib@innerbib\@empty
\bibitem [{\citenamefont {Fisch}(1978{\natexlab{a}})}]{Fisch1978}%
  \BibitemOpen
  \bibfield  {author} {\bibinfo {author} {\bibfnamefont {N.~J.}\ \bibnamefont
  {Fisch}},\ }\bibfield  {title} {\enquote {\bibinfo {title} {{Confining a
  tokamak plasma with rf-driven currents}},}\ }\href {\doibase
  https://doi.org/10.1103/PhysRevLett.41.873} {\bibfield  {journal} {\bibinfo
  {journal} {Physical Review Letters}\ }\textbf {\bibinfo {volume} {41}},\
  \bibinfo {pages} {873} (\bibinfo {year} {1978}{\natexlab{a}})}\BibitemShut
  {NoStop}%
\bibitem [{\citenamefont {Fisch}\ and\ \citenamefont
  {Rax}(1992{\natexlab{a}})}]{Fisch1992}%
  \BibitemOpen
  \bibfield  {author} {\bibinfo {author} {\bibfnamefont {N.~J.}\ \bibnamefont
  {Fisch}}\ and\ \bibinfo {author} {\bibfnamefont {J.-M.}\ \bibnamefont
  {Rax}},\ }\bibfield  {title} {\enquote {\bibinfo {title} {{Interaction of
  energetic alpha particles with intense lower hybrid waves}},}\ }\href@noop {}
  {\bibfield  {journal} {\bibinfo  {journal} {Physical Review Letters}\
  }\textbf {\bibinfo {volume} {69}},\ \bibinfo {pages} {612} (\bibinfo {year}
  {1992}{\natexlab{a}})}\BibitemShut {NoStop}%
\bibitem [{\citenamefont {Fisch}\ and\ \citenamefont
  {Rax}(1992{\natexlab{b}})}]{fisch1992current}%
  \BibitemOpen
  \bibfield  {author} {\bibinfo {author} {\bibfnamefont {N.~J.}\ \bibnamefont
  {Fisch}}\ and\ \bibinfo {author} {\bibfnamefont {J.~M.}\ \bibnamefont
  {Rax}},\ }\bibfield  {title} {\enquote {\bibinfo {title} {{Current Drive by
  Lower Hybrid Waves in the Presence of Energetic Alpha Particles}},}\
  }\href@noop {} {\bibfield  {journal} {\bibinfo  {journal} {Nuclear Fusion}\
  }\textbf {\bibinfo {volume} {32}},\ \bibinfo {pages} {549} (\bibinfo {year}
  {1992}{\natexlab{b}})}\BibitemShut {NoStop}%
\bibitem [{\citenamefont {Heikkinen}\ and\ \citenamefont
  {Sipila}(1996)}]{Heikkinen1996}%
  \BibitemOpen
  \bibfield  {author} {\bibinfo {author} {\bibfnamefont {J.~A.}\ \bibnamefont
  {Heikkinen}}\ and\ \bibinfo {author} {\bibfnamefont {S.~K.}\ \bibnamefont
  {Sipila}},\ }\bibfield  {title} {\enquote {\bibinfo {title} {{Current driven
  by lower hybrid heating of thermonuclear alpha particles in Tokamak
  reactors}},}\ }\href {\doibase 10.1088/0029-5515/36/10/I07} {\bibfield
  {journal} {\bibinfo  {journal} {Nuclear Fusion}\ }\textbf {\bibinfo {volume}
  {36}},\ \bibinfo {pages} {1345--1355} (\bibinfo {year} {1996})}\BibitemShut
  {NoStop}%
\bibitem [{\citenamefont {Ochs}, \citenamefont {Bertelli},\ and\ \citenamefont
  {Fisch}(2015{\natexlab{a}})}]{ochs2015coupling}%
  \BibitemOpen
  \bibfield  {author} {\bibinfo {author} {\bibfnamefont {I.~E.}\ \bibnamefont
  {Ochs}}, \bibinfo {author} {\bibfnamefont {N.}~\bibnamefont {Bertelli}}, \
  and\ \bibinfo {author} {\bibfnamefont {N.~J.}\ \bibnamefont {Fisch}},\
  }\bibfield  {title} {\enquote {\bibinfo {title} {{Coupling of alpha
  channeling to parallel wavenumber upshift in lower hybrid current drive}},}\
  }\href {\doibase 10.1063/1.4928903} {\bibfield  {journal} {\bibinfo
  {journal} {Physics of Plasmas}\ }\textbf {\bibinfo {volume} {22}},\ \bibinfo
  {pages} {1--4} (\bibinfo {year} {2015}{\natexlab{a}})}\BibitemShut {NoStop}%
\bibitem [{\citenamefont {Ochs}, \citenamefont {Bertelli},\ and\ \citenamefont
  {Fisch}(2015{\natexlab{b}})}]{ochs2015alpha}%
  \BibitemOpen
  \bibfield  {author} {\bibinfo {author} {\bibfnamefont {I.~E.}\ \bibnamefont
  {Ochs}}, \bibinfo {author} {\bibfnamefont {N.}~\bibnamefont {Bertelli}}, \
  and\ \bibinfo {author} {\bibfnamefont {N.~J.}\ \bibnamefont {Fisch}},\
  }\bibfield  {title} {\enquote {\bibinfo {title} {{Alpha channeling with
  high-field launch of lower hybrid waves}},}\ }\href {\doibase
  10.1063/1.4935123} {\bibfield  {journal} {\bibinfo  {journal} {Physics of
  Plasmas}\ }\textbf {\bibinfo {volume} {22}},\ \bibinfo {pages} {112103}
  (\bibinfo {year} {2015}{\natexlab{b}})}\BibitemShut {NoStop}%
\bibitem [{\citenamefont {Valeo}\ and\ \citenamefont
  {Fisch}(1994)}]{Valeo1994}%
  \BibitemOpen
  \bibfield  {author} {\bibinfo {author} {\bibfnamefont {E.~J.}\ \bibnamefont
  {Valeo}}\ and\ \bibinfo {author} {\bibfnamefont {N.~J.}\ \bibnamefont
  {Fisch}},\ }\bibfield  {title} {\enquote {\bibinfo {title} {{Excitation of
  Large-kTheta Ion-Bernstein Waves in Tokamaks}},}\ }\href@noop {} {\bibfield
  {journal} {\bibinfo  {journal} {Phys. Rev. Lett.}\ }\textbf {\bibinfo
  {volume} {73}},\ \bibinfo {pages} {3536} (\bibinfo {year}
  {1994})}\BibitemShut {NoStop}%
\bibitem [{\citenamefont {Fisch}(1995)}]{fisch1995ibw}%
  \BibitemOpen
  \bibfield  {author} {\bibinfo {author} {\bibfnamefont {N.~J.}\ \bibnamefont
  {Fisch}},\ }\bibfield  {title} {\enquote {\bibinfo {title} {{Alpha power
  channeling using ion-{Bernstein} waves}},}\ }\href@noop {} {\bibfield
  {journal} {\bibinfo  {journal} {Phys.$\sim$Plasmas}\ }\textbf {\bibinfo
  {volume} {2}},\ \bibinfo {pages} {2375} (\bibinfo {year} {1995})}\BibitemShut
  {NoStop}%
\bibitem [{\citenamefont {Fisch}\ and\ \citenamefont
  {Herrmann}(1995)}]{Fisch1995a}%
  \BibitemOpen
  \bibfield  {author} {\bibinfo {author} {\bibfnamefont {N.~J.}\ \bibnamefont
  {Fisch}}\ and\ \bibinfo {author} {\bibfnamefont {M.~C.}\ \bibnamefont
  {Herrmann}},\ }\bibfield  {title} {\enquote {\bibinfo {title} {{Alpha power
  channelling with two waves}},}\ }\href@noop {} {\bibfield  {journal}
  {\bibinfo  {journal} {Nuclear Fusion}\ }\textbf {\bibinfo {volume} {35}},\
  \bibinfo {pages} {1753} (\bibinfo {year} {1995})}\BibitemShut {NoStop}%
\bibitem [{\citenamefont {Heikkinen}\ and\ \citenamefont
  {Sipil{\"{a}}}(1995)}]{Heikkinen1995}%
  \BibitemOpen
  \bibfield  {author} {\bibinfo {author} {\bibfnamefont {J.~A.}\ \bibnamefont
  {Heikkinen}}\ and\ \bibinfo {author} {\bibfnamefont {S.~K.}\ \bibnamefont
  {Sipil{\"{a}}}},\ }\bibfield  {title} {\enquote {\bibinfo {title} {{Power
  transfer and current generation of fast ions with large-k theta waves in
  tokamak plasmas}},}\ }\href {\doibase 10.1063/1.871072} {\bibfield  {journal}
  {\bibinfo  {journal} {Physics of Plasmas}\ }\textbf {\bibinfo {volume} {2}},\
  \bibinfo {pages} {3724--3733} (\bibinfo {year} {1995})}\BibitemShut {NoStop}%
\bibitem [{\citenamefont {Marchenko}(1998)}]{Marchenko1998}%
  \BibitemOpen
  \bibfield  {author} {\bibinfo {author} {\bibfnamefont {V.~S.}\ \bibnamefont
  {Marchenko}},\ }\bibfield  {title} {\enquote {\bibinfo {title} {{Interaction
  of large amplitude ion Bernstein waves with hot ions in tokamaks}},}\ }\href
  {\doibase 10.1088/0029-5515/38/10/101} {\bibfield  {journal} {\bibinfo
  {journal} {Nuclear Fusion}\ }\textbf {\bibinfo {volume} {38}},\ \bibinfo
  {pages} {1427--1430} (\bibinfo {year} {1998})}\BibitemShut {NoStop}%
\bibitem [{\citenamefont {Kuley}, \citenamefont {Liu},\ and\ \citenamefont
  {Tripathi}(2011)}]{Kuley2011}%
  \BibitemOpen
  \bibfield  {author} {\bibinfo {author} {\bibfnamefont {A.}~\bibnamefont
  {Kuley}}, \bibinfo {author} {\bibfnamefont {C.~S.}\ \bibnamefont {Liu}}, \
  and\ \bibinfo {author} {\bibfnamefont {V.~K.}\ \bibnamefont {Tripathi}},\
  }\bibfield  {title} {\enquote {\bibinfo {title} {{Energy channeling due to
  energetic-ion-driven instabilities in tokamak}},}\ }\href {\doibase
  10.1063/1.3561831} {\bibfield  {journal} {\bibinfo  {journal} {Physics of
  Plasmas}\ }\textbf {\bibinfo {volume} {18}} (\bibinfo {year} {2011}),\
  10.1063/1.3561831}\BibitemShut {NoStop}%
\bibitem [{\citenamefont {Sasaki}, \citenamefont {Itoh},\ and\ \citenamefont
  {Itoh}(2011)}]{Sasaki2011}%
  \BibitemOpen
  \bibfield  {author} {\bibinfo {author} {\bibfnamefont {M.}~\bibnamefont
  {Sasaki}}, \bibinfo {author} {\bibfnamefont {K.}~\bibnamefont {Itoh}}, \ and\
  \bibinfo {author} {\bibfnamefont {S.~I.}\ \bibnamefont {Itoh}},\ }\bibfield
  {title} {\enquote {\bibinfo {title} {{Energy channeling from energetic
  particles to bulk ions via beam-driven geodesic acoustic modes - GAM
  channeling}},}\ }\href {\doibase 10.1088/0741-3335/53/8/085017} {\bibfield
  {journal} {\bibinfo  {journal} {Plasma Physics and Controlled Fusion}\
  }\textbf {\bibinfo {volume} {53}} (\bibinfo {year} {2011}),\
  10.1088/0741-3335/53/8/085017}\BibitemShut {NoStop}%
\bibitem [{\citenamefont {Chen}\ and\ \citenamefont {Zonca}(2016)}]{Chen2016a}%
  \BibitemOpen
  \bibfield  {author} {\bibinfo {author} {\bibfnamefont {L.}~\bibnamefont
  {Chen}}\ and\ \bibinfo {author} {\bibfnamefont {F.}~\bibnamefont {Zonca}},\
  }\bibfield  {title} {\enquote {\bibinfo {title} {{Physics of Alfven waves and
  energetic particles in burning plasmas}},}\ }\href@noop {} {\bibfield
  {journal} {\bibinfo  {journal} {Rev. Mod. Phys.}\ }\textbf {\bibinfo {volume}
  {88}},\ \bibinfo {pages} {15008} (\bibinfo {year} {2016})}\BibitemShut
  {NoStop}%
\bibitem [{\citenamefont {Gorelenkov}(2016)}]{Gorelenkov2016}%
  \BibitemOpen
  \bibfield  {author} {\bibinfo {author} {\bibfnamefont {N.~N.}\ \bibnamefont
  {Gorelenkov}},\ }\bibfield  {title} {\enquote {\bibinfo {title} {{Energetic
  particle-driven compressional Alfven eigenmodes and prospects for ion
  cyclotron emission studies in fusion plasmas}},}\ }\href@noop {} {\bibfield
  {journal} {\bibinfo  {journal} {New J. Phys.}\ }\textbf {\bibinfo {volume}
  {18}},\ \bibinfo {pages} {105010} (\bibinfo {year} {2016})}\BibitemShut
  {NoStop}%
\bibitem [{\citenamefont {Cook}, \citenamefont {Dendy},\ and\ \citenamefont
  {Chapman}(2017)}]{Cook2017}%
  \BibitemOpen
  \bibfield  {author} {\bibinfo {author} {\bibfnamefont {J.~W.~S.}\
  \bibnamefont {Cook}}, \bibinfo {author} {\bibfnamefont {R.~O.}\ \bibnamefont
  {Dendy}}, \ and\ \bibinfo {author} {\bibfnamefont {S.~C.}\ \bibnamefont
  {Chapman}},\ }\bibfield  {title} {\enquote {\bibinfo {title} {{Stimulated
  Emission of Fast Alfven Waves within Magnetically Confined Fusion
  Plasmas}},}\ }\href@noop {} {\bibfield  {journal} {\bibinfo  {journal} {Phys.
  Rev. Lett.}\ }\textbf {\bibinfo {volume} {118}},\ \bibinfo {pages} {185001}
  (\bibinfo {year} {2017})}\BibitemShut {NoStop}%
\bibitem [{\citenamefont {Castaldo}, \citenamefont {Cardinali},\ and\
  \citenamefont {Napoli}(2019)}]{Castaldo2019}%
  \BibitemOpen
  \bibfield  {author} {\bibinfo {author} {\bibfnamefont {C.}~\bibnamefont
  {Castaldo}}, \bibinfo {author} {\bibfnamefont {A.}~\bibnamefont {Cardinali}},
  \ and\ \bibinfo {author} {\bibfnamefont {F.}~\bibnamefont {Napoli}},\
  }\bibfield  {title} {\enquote {\bibinfo {title} {{Nonlinear inverse Landau
  damping of ion Bernstein waves on alpha particles}},}\ }\href {\doibase
  10.1088/1361-6587/ab2226} {\bibfield  {journal} {\bibinfo  {journal} {Plasma
  Physics and Controlled Fusion}\ }\textbf {\bibinfo {volume} {61}},\ \bibinfo
  {pages} {084007} (\bibinfo {year} {2019})}\BibitemShut {NoStop}%
\bibitem [{\citenamefont {Romanelli}\ and\ \citenamefont
  {Cardinali}(2020)}]{Romanelli2020}%
  \BibitemOpen
  \bibfield  {author} {\bibinfo {author} {\bibfnamefont {F.}~\bibnamefont
  {Romanelli}}\ and\ \bibinfo {author} {\bibfnamefont {A.}~\bibnamefont
  {Cardinali}},\ }\bibfield  {title} {\enquote {\bibinfo {title} {{On the
  interaction of ion Bernstein waves with alpha particles}},}\ }\href {\doibase
  10.1088/1741-4326/ab6c78} {\bibfield  {journal} {\bibinfo  {journal} {Nuclear
  Fusion}\ }\textbf {\bibinfo {volume} {60}},\ \bibinfo {pages} {036025}
  (\bibinfo {year} {2020})}\BibitemShut {NoStop}%
\bibitem [{\citenamefont {Herrmann}\ and\ \citenamefont
  {Fisch}(1997)}]{Herrmann1997}%
  \BibitemOpen
  \bibfield  {author} {\bibinfo {author} {\bibfnamefont {M.~C.}\ \bibnamefont
  {Herrmann}}\ and\ \bibinfo {author} {\bibfnamefont {N.~J.}\ \bibnamefont
  {Fisch}},\ }\bibfield  {title} {\enquote {\bibinfo {title} {{Cooling
  Energetic alpha particles in a Tokamak with Waves}},}\ }\href@noop {}
  {\bibfield  {journal} {\bibinfo  {journal} {Physical Review Letters}\
  }\textbf {\bibinfo {volume} {79}},\ \bibinfo {pages} {1495} (\bibinfo {year}
  {1997})}\BibitemShut {NoStop}%
\bibitem [{\citenamefont {Cianfrani}\ and\ \citenamefont
  {Romanelli}(2018)}]{Cianfrani2018}%
  \BibitemOpen
  \bibfield  {author} {\bibinfo {author} {\bibfnamefont {F.}~\bibnamefont
  {Cianfrani}}\ and\ \bibinfo {author} {\bibfnamefont {F.}~\bibnamefont
  {Romanelli}},\ }\bibfield  {title} {\enquote {\bibinfo {title} {{On the
  optimal conditions for alpha-channelling}},}\ }\href {\doibase
  10.1088/1741-4326/aac128} {\bibfield  {journal} {\bibinfo  {journal} {Nucl.
  Fusion}\ }\textbf {\bibinfo {volume} {58}},\ \bibinfo {pages} {76013}
  (\bibinfo {year} {2018})}\BibitemShut {NoStop}%
\bibitem [{\citenamefont {Cianfrani}\ and\ \citenamefont
  {Romanelli}(2019)}]{Cianfrani2019}%
  \BibitemOpen
  \bibfield  {author} {\bibinfo {author} {\bibfnamefont {F.}~\bibnamefont
  {Cianfrani}}\ and\ \bibinfo {author} {\bibfnamefont {F.}~\bibnamefont
  {Romanelli}},\ }\bibfield  {title} {\enquote {\bibinfo {title} {{On the
  optimal conditions for alpha channelling in tokamaks}},}\ }\href {\doibase
  10.1088/1741-4326/ab3018} {\bibfield  {journal} {\bibinfo  {journal} {Nuclear
  Fusion}\ }\textbf {\bibinfo {volume} {59}},\ \bibinfo {pages} {106005}
  (\bibinfo {year} {2019})}\BibitemShut {NoStop}%
\bibitem [{\citenamefont {White}\ \emph {et~al.}(2021)\citenamefont {White},
  \citenamefont {Romanelli}, \citenamefont {Cianfrani},\ and\ \citenamefont
  {Valeo}}]{White2021}%
  \BibitemOpen
  \bibfield  {author} {\bibinfo {author} {\bibfnamefont {R.}~\bibnamefont
  {White}}, \bibinfo {author} {\bibfnamefont {F.}~\bibnamefont {Romanelli}},
  \bibinfo {author} {\bibfnamefont {F.}~\bibnamefont {Cianfrani}}, \ and\
  \bibinfo {author} {\bibfnamefont {E.}~\bibnamefont {Valeo}},\ }\bibfield
  {title} {\enquote {\bibinfo {title} {{Alpha particle channeling in ITER}},}\
  }\href@noop {} {\bibfield  {journal} {\bibinfo  {journal} {Phys. Plasmas}\
  }\textbf {\bibinfo {volume} {28}},\ \bibinfo {pages} {12503} (\bibinfo {year}
  {2021})}\BibitemShut {NoStop}%
\bibitem [{\citenamefont {Fetterman}\ and\ \citenamefont
  {Fisch}(2008)}]{fetterman2008alpha}%
  \BibitemOpen
  \bibfield  {author} {\bibinfo {author} {\bibfnamefont {A.~J.}\ \bibnamefont
  {Fetterman}}\ and\ \bibinfo {author} {\bibfnamefont {N.~J.}\ \bibnamefont
  {Fisch}},\ }\bibfield  {title} {\enquote {\bibinfo {title} {{Alpha channeling
  in a rotating plasma}},}\ }\href@noop {} {\bibfield  {journal} {\bibinfo
  {journal} {Physical Review Letters}\ }\textbf {\bibinfo {volume} {101}},\
  \bibinfo {pages} {205003} (\bibinfo {year} {2008})}\BibitemShut {NoStop}%
\bibitem [{\citenamefont {Fetterman}(2012)}]{fetterman2012wave}%
  \BibitemOpen
  \bibfield  {author} {\bibinfo {author} {\bibfnamefont {A.~J.}\ \bibnamefont
  {Fetterman}},\ }\emph {\bibinfo {title} {{Wave-driven rotation and mass
  separation in rotating magnetic mirrors}}},\ \href@noop {} {Ph.D. thesis},\
  \bibinfo  {school} {Princeton University} (\bibinfo {year}
  {2012})\BibitemShut {NoStop}%
\bibitem [{\citenamefont {Maggs}, \citenamefont {Carter},\ and\ \citenamefont
  {Taylor}(2007)}]{maggs2007transition}%
  \BibitemOpen
  \bibfield  {author} {\bibinfo {author} {\bibfnamefont {J.~E.}\ \bibnamefont
  {Maggs}}, \bibinfo {author} {\bibfnamefont {T.~A.}\ \bibnamefont {Carter}}, \
  and\ \bibinfo {author} {\bibfnamefont {R.~J.}\ \bibnamefont {Taylor}},\
  }\bibfield  {title} {\enquote {\bibinfo {title} {{Transition from Bohm to
  classical diffusion due to edge rotation of a cylindrical plasma}},}\
  }\href@noop {} {\bibfield  {journal} {\bibinfo  {journal} {Physics of
  Plasmas}\ }\textbf {\bibinfo {volume} {14}},\ \bibinfo {pages} {052507}
  (\bibinfo {year} {2007})}\BibitemShut {NoStop}%
\bibitem [{\citenamefont {Burrell}(2020)}]{Burrell2020}%
  \BibitemOpen
  \bibfield  {author} {\bibinfo {author} {\bibfnamefont {K.~H.}\ \bibnamefont
  {Burrell}},\ }\bibfield  {title} {\enquote {\bibinfo {title} {{Role of
  sheared ExB flow in self-organized, improved confinement states in magnetized
  plasmas}},}\ }\href {\doibase 10.1063/1.5142734} {\bibfield  {journal}
  {\bibinfo  {journal} {Physics of Plasmas}\ }\textbf {\bibinfo {volume} {27}}
  (\bibinfo {year} {2020}),\ 10.1063/1.5142734}\BibitemShut {NoStop}%
\bibitem [{\citenamefont {Shumlak}\ and\ \citenamefont
  {Hartman}(1995)}]{Shumlak1995}%
  \BibitemOpen
  \bibfield  {author} {\bibinfo {author} {\bibfnamefont {U.}~\bibnamefont
  {Shumlak}}\ and\ \bibinfo {author} {\bibfnamefont {C.~W.}\ \bibnamefont
  {Hartman}},\ }\bibfield  {title} {\enquote {\bibinfo {title} {{Sheared Flow
  Stabilization of the m = 1 Kink Mode in Z Pinches}},}\ }\href@noop {}
  {\bibfield  {journal} {\bibinfo  {journal} {Physical Review Letters}\
  }\textbf {\bibinfo {volume} {75}},\ \bibinfo {pages} {3285} (\bibinfo {year}
  {1995})}\BibitemShut {NoStop}%
\bibitem [{\citenamefont {Huang}\ and\ \citenamefont
  {Hassam}(2001)}]{Huang2001}%
  \BibitemOpen
  \bibfield  {author} {\bibinfo {author} {\bibfnamefont {Y.~M.}\ \bibnamefont
  {Huang}}\ and\ \bibinfo {author} {\bibfnamefont {A.~B.}\ \bibnamefont
  {Hassam}},\ }\bibfield  {title} {\enquote {\bibinfo {title} {{Velocity shear
  stabilization of centrifugally confined plasma}},}\ }\href {\doibase
  10.1103/PhysRevLett.87.235002} {\bibfield  {journal} {\bibinfo  {journal}
  {Physical Review Letters}\ }\textbf {\bibinfo {volume} {87}},\ \bibinfo
  {pages} {235002--1--235002--4} (\bibinfo {year} {2001})}\BibitemShut
  {NoStop}%
\bibitem [{\citenamefont {Kolmes}\ \emph {et~al.}(2021)\citenamefont {Kolmes},
  \citenamefont {Ochs}, \citenamefont {Mlodik},\ and\ \citenamefont
  {Fisch}}]{Kolmes2021a}%
  \BibitemOpen
  \bibfield  {author} {\bibinfo {author} {\bibfnamefont {E.~J.}\ \bibnamefont
  {Kolmes}}, \bibinfo {author} {\bibfnamefont {I.~E.}\ \bibnamefont {Ochs}},
  \bibinfo {author} {\bibfnamefont {M.~E.}\ \bibnamefont {Mlodik}}, \ and\
  \bibinfo {author} {\bibfnamefont {N.~J.}\ \bibnamefont {Fisch}},\ }\bibfield
  {title} {\enquote {\bibinfo {title} {{Natural Hot Ion Modes in a Rotating
  Plasma}},}\ }\href@noop {} {\bibfield  {journal} {\bibinfo  {journal} {Arxiv
  Preprint}\ } (\bibinfo {year} {2021})},\ \Eprint
  {http://arxiv.org/abs/2101.00409} {arXiv:2101.00409} \BibitemShut {NoStop}%
\bibitem [{\citenamefont {Krall}\ and\ \citenamefont
  {Trivelpiece}(1973)}]{kralltrivelpiece}%
  \BibitemOpen
  \bibfield  {author} {\bibinfo {author} {\bibfnamefont {N.~A.}\ \bibnamefont
  {Krall}}\ and\ \bibinfo {author} {\bibfnamefont {A.~W.}\ \bibnamefont
  {Trivelpiece}},\ }\href@noop {} {\emph {\bibinfo {title} {{Principles of
  Plasma Physics}}}}\ (\bibinfo  {publisher} {McGraw-Hill},\ \bibinfo {year}
  {1973})\BibitemShut {NoStop}%
\bibitem [{\citenamefont {Davidson}\ and\ \citenamefont
  {Scherer}(1972)}]{davidson1973methods}%
  \BibitemOpen
  \bibfield  {author} {\bibinfo {author} {\bibfnamefont {R.~C.}\ \bibnamefont
  {Davidson}}\ and\ \bibinfo {author} {\bibfnamefont {J.~E.}\ \bibnamefont
  {Scherer}},\ }\href@noop {} {\emph {\bibinfo {title} {IEEE Transactions on
  Plasma Science}}}\ (\bibinfo  {publisher} {Academic Press},\ \bibinfo {year}
  {1972})\ p.~\bibinfo {pages} {58}\BibitemShut {NoStop}%
\bibitem [{\citenamefont {Stix}(1992)}]{stix1992waves}%
  \BibitemOpen
  \bibfield  {author} {\bibinfo {author} {\bibfnamefont {T.~H.}\ \bibnamefont
  {Stix}},\ }\href@noop {} {\emph {\bibinfo {title} {{Waves in Plasmas}}}}\
  (\bibinfo  {publisher} {AIP Press},\ \bibinfo {year} {1992})\BibitemShut
  {NoStop}%
\bibitem [{\citenamefont {Yamamoto}\ \emph {et~al.}(1980)\citenamefont
  {Yamamoto}, \citenamefont {Imai}, \citenamefont {Shimada}, \citenamefont
  {Suzuki}, \citenamefont {Maeno}, \citenamefont {Konoshima}, \citenamefont
  {Fujii}, \citenamefont {Uehara}, \citenamefont {Nagashima}, \citenamefont
  {Funahashi},\ and\ \citenamefont {Fujisawa}}]{Yamamoto1980}%
  \BibitemOpen
  \bibfield  {author} {\bibinfo {author} {\bibfnamefont {T.}~\bibnamefont
  {Yamamoto}}, \bibinfo {author} {\bibfnamefont {T.}~\bibnamefont {Imai}},
  \bibinfo {author} {\bibfnamefont {M.}~\bibnamefont {Shimada}}, \bibinfo
  {author} {\bibfnamefont {N.}~\bibnamefont {Suzuki}}, \bibinfo {author}
  {\bibfnamefont {M.}~\bibnamefont {Maeno}}, \bibinfo {author} {\bibfnamefont
  {S.}~\bibnamefont {Konoshima}}, \bibinfo {author} {\bibfnamefont
  {T.}~\bibnamefont {Fujii}}, \bibinfo {author} {\bibfnamefont
  {K.}~\bibnamefont {Uehara}}, \bibinfo {author} {\bibfnamefont
  {T.}~\bibnamefont {Nagashima}}, \bibinfo {author} {\bibfnamefont
  {A.}~\bibnamefont {Funahashi}}, \ and\ \bibinfo {author} {\bibfnamefont
  {N.}~\bibnamefont {Fujisawa}},\ }\bibfield  {title} {\enquote {\bibinfo
  {title} {{Experimental observation of the rf-driven current by the
  lower-Hybrid wave in a tokamak}},}\ }\href {\doibase
  10.1103/PhysRevLett.45.716} {\bibfield  {journal} {\bibinfo  {journal}
  {Physical Review Letters}\ }\textbf {\bibinfo {volume} {45}},\ \bibinfo
  {pages} {716--719} (\bibinfo {year} {1980})}\BibitemShut {NoStop}%
\bibitem [{\citenamefont {Wong}, \citenamefont {Horton},\ and\ \citenamefont
  {Ono}(1980)}]{Wong1980}%
  \BibitemOpen
  \bibfield  {author} {\bibinfo {author} {\bibfnamefont {K.-L.}\ \bibnamefont
  {Wong}}, \bibinfo {author} {\bibfnamefont {R.}~\bibnamefont {Horton}}, \ and\
  \bibinfo {author} {\bibfnamefont {M.}~\bibnamefont {Ono}},\ }\bibfield
  {title} {\enquote {\bibinfo {title} {{Current generation by lower hybrid
  waves in the ACT-1 toroidal device}},}\ }\href@noop {} {\bibfield  {journal}
  {\bibinfo  {journal} {Physical Review Letters}\ }\textbf {\bibinfo {volume}
  {45}},\ \bibinfo {pages} {117--121} (\bibinfo {year} {1980})}\BibitemShut
  {NoStop}%
\bibitem [{\citenamefont {Kojima}, \citenamefont {Takamura},\ and\
  \citenamefont {Okuda}(1981)}]{Kojima1981}%
  \BibitemOpen
  \bibfield  {author} {\bibinfo {author} {\bibfnamefont {T.}~\bibnamefont
  {Kojima}}, \bibinfo {author} {\bibfnamefont {S.}~\bibnamefont {Takamura}}, \
  and\ \bibinfo {author} {\bibfnamefont {T.}~\bibnamefont {Okuda}},\ }\bibfield
   {title} {\enquote {\bibinfo {title} {{Current generation due to a travelling
  lower hybrid wave excited by helical antennas}},}\ }\href {\doibase
  10.1016/0375-9601(81)90878-1} {\bibfield  {journal} {\bibinfo  {journal}
  {Physics Letters A}\ }\textbf {\bibinfo {volume} {83}},\ \bibinfo {pages}
  {172--174} (\bibinfo {year} {1981})}\BibitemShut {NoStop}%
\bibitem [{\citenamefont {Bernabei}\ \emph {et~al.}(1982)\citenamefont
  {Bernabei}, \citenamefont {Daughney}, \citenamefont {Efthimion},
  \citenamefont {Hooke}, \citenamefont {Hosea}, \citenamefont {Jobes},
  \citenamefont {Martin}, \citenamefont {Mazzucato}, \citenamefont {Meservey},
  \citenamefont {Motley}, \citenamefont {Stevens}, \citenamefont {{Von
  Goeler}},\ and\ \citenamefont {Wilson}}]{Bernabei1982}%
  \BibitemOpen
  \bibfield  {author} {\bibinfo {author} {\bibfnamefont {S.}~\bibnamefont
  {Bernabei}}, \bibinfo {author} {\bibfnamefont {C.}~\bibnamefont {Daughney}},
  \bibinfo {author} {\bibfnamefont {P.}~\bibnamefont {Efthimion}}, \bibinfo
  {author} {\bibfnamefont {W.}~\bibnamefont {Hooke}}, \bibinfo {author}
  {\bibfnamefont {J.}~\bibnamefont {Hosea}}, \bibinfo {author} {\bibfnamefont
  {F.}~\bibnamefont {Jobes}}, \bibinfo {author} {\bibfnamefont
  {A.}~\bibnamefont {Martin}}, \bibinfo {author} {\bibfnamefont
  {E.}~\bibnamefont {Mazzucato}}, \bibinfo {author} {\bibfnamefont
  {E.}~\bibnamefont {Meservey}}, \bibinfo {author} {\bibfnamefont
  {R.}~\bibnamefont {Motley}}, \bibinfo {author} {\bibfnamefont
  {J.}~\bibnamefont {Stevens}}, \bibinfo {author} {\bibfnamefont
  {S.}~\bibnamefont {{Von Goeler}}}, \ and\ \bibinfo {author} {\bibfnamefont
  {R.}~\bibnamefont {Wilson}},\ }\bibfield  {title} {\enquote {\bibinfo {title}
  {{Lower-Hybrid Current Drive in the {PLT} Tokamak}},}\ }\href@noop {}
  {\bibfield  {journal} {\bibinfo  {journal} {Physical Review Letters}\
  }\textbf {\bibinfo {volume} {49}},\ \bibinfo {pages} {1255} (\bibinfo {year}
  {1982})}\BibitemShut {NoStop}%
\bibitem [{\citenamefont {Porkolab}\ \emph {et~al.}(1984)\citenamefont
  {Porkolab}, \citenamefont {Schuss}, \citenamefont {Lloyd}, \citenamefont
  {Takase}, \citenamefont {Texter}, \citenamefont {Bonoli}, \citenamefont
  {Fiore}, \citenamefont {Gandy}, \citenamefont {Gwinn}, \citenamefont
  {Lipschultz},\ and\ \citenamefont {Others}}]{Porkolab1984}%
  \BibitemOpen
  \bibfield  {author} {\bibinfo {author} {\bibfnamefont {M.}~\bibnamefont
  {Porkolab}}, \bibinfo {author} {\bibfnamefont {J.~J.}\ \bibnamefont
  {Schuss}}, \bibinfo {author} {\bibfnamefont {B.}~\bibnamefont {Lloyd}},
  \bibinfo {author} {\bibfnamefont {Y.}~\bibnamefont {Takase}}, \bibinfo
  {author} {\bibfnamefont {S.}~\bibnamefont {Texter}}, \bibinfo {author}
  {\bibfnamefont {P.}~\bibnamefont {Bonoli}}, \bibinfo {author} {\bibfnamefont
  {C.}~\bibnamefont {Fiore}}, \bibinfo {author} {\bibfnamefont
  {R.}~\bibnamefont {Gandy}}, \bibinfo {author} {\bibfnamefont
  {D.}~\bibnamefont {Gwinn}}, \bibinfo {author} {\bibfnamefont
  {B.}~\bibnamefont {Lipschultz}}, \ and\ \bibinfo {author} {\bibnamefont
  {Others}},\ }\bibfield  {title} {\enquote {\bibinfo {title} {{Observation of
  lower-hybrid current drive at high densities in the {A}lcator {C}
  tokamak}},}\ }\href@noop {} {\bibfield  {journal} {\bibinfo  {journal}
  {Physical Review Letters}\ }\textbf {\bibinfo {volume} {53}},\ \bibinfo
  {pages} {450} (\bibinfo {year} {1984})}\BibitemShut {NoStop}%
\bibitem [{\citenamefont {Ekedahl}\ \emph {et~al.}(1998)\citenamefont
  {Ekedahl}, \citenamefont {Baranov}, \citenamefont {Dobbing}, \citenamefont
  {Fischer}, \citenamefont {Gormezano}, \citenamefont {Jones}, \citenamefont
  {Lennholm}, \citenamefont {Parail}, \citenamefont {Rimini}, \citenamefont
  {Romero}, \citenamefont {Schild}, \citenamefont {Sips}, \citenamefont
  {S{\"{o}}ldner},\ and\ \citenamefont {Tubbing}}]{Ekedahl1998}%
  \BibitemOpen
  \bibfield  {author} {\bibinfo {author} {\bibfnamefont {A.}~\bibnamefont
  {Ekedahl}}, \bibinfo {author} {\bibfnamefont {Y.~F.}\ \bibnamefont
  {Baranov}}, \bibinfo {author} {\bibfnamefont {J.~A.}\ \bibnamefont
  {Dobbing}}, \bibinfo {author} {\bibfnamefont {B.}~\bibnamefont {Fischer}},
  \bibinfo {author} {\bibfnamefont {C.}~\bibnamefont {Gormezano}}, \bibinfo
  {author} {\bibfnamefont {T.~T.}\ \bibnamefont {Jones}}, \bibinfo {author}
  {\bibfnamefont {M.}~\bibnamefont {Lennholm}}, \bibinfo {author}
  {\bibfnamefont {V.~V.}\ \bibnamefont {Parail}}, \bibinfo {author}
  {\bibfnamefont {F.~G.}\ \bibnamefont {Rimini}}, \bibinfo {author}
  {\bibfnamefont {J.~A.}\ \bibnamefont {Romero}}, \bibinfo {author}
  {\bibfnamefont {P.}~\bibnamefont {Schild}}, \bibinfo {author} {\bibfnamefont
  {A.~C.}\ \bibnamefont {Sips}}, \bibinfo {author} {\bibfnamefont {F.~X.}\
  \bibnamefont {S{\"{o}}ldner}}, \ and\ \bibinfo {author} {\bibfnamefont
  {B.~J.}\ \bibnamefont {Tubbing}},\ }\bibfield  {title} {\enquote {\bibinfo
  {title} {{Profile control experiments in jet using off-axis lower hybrid
  current drive}},}\ }\href {\doibase 10.1088/0029-5515/38/9/314} {\bibfield
  {journal} {\bibinfo  {journal} {Nuclear Fusion}\ }\textbf {\bibinfo {volume}
  {38}},\ \bibinfo {pages} {1397--1407} (\bibinfo {year} {1998})}\BibitemShut
  {NoStop}%
\bibitem [{\citenamefont {Kato}(1980)}]{kato1980electrostatic}%
  \BibitemOpen
  \bibfield  {author} {\bibinfo {author} {\bibfnamefont {K.}~\bibnamefont
  {Kato}},\ }\bibfield  {title} {\enquote {\bibinfo {title} {{Theory of Current
  Generation by Electrostatic Traveling Waves in Collisionless Magnetized
  Plasmas}},}\ }\href {\doibase 10.1103/PhysRevLett.44.779} {\bibfield
  {journal} {\bibinfo  {journal} {Phys. Rev. Lett.}\ }\textbf {\bibinfo
  {volume} {44}},\ \bibinfo {pages} {779--781} (\bibinfo {year}
  {1980})}\BibitemShut {NoStop}%
\bibitem [{\citenamefont {Ochs}\ and\ \citenamefont
  {Fisch}(2020{\natexlab{a}})}]{Ochs2020}%
  \BibitemOpen
  \bibfield  {author} {\bibinfo {author} {\bibfnamefont {I.~E.}\ \bibnamefont
  {Ochs}}\ and\ \bibinfo {author} {\bibfnamefont {N.~J.}\ \bibnamefont
  {Fisch}},\ }\bibfield  {title} {\enquote {\bibinfo {title}
  {{Momentum-exchange current drive by electrostatic waves in an unmagnetized
  collisionless plasma}},}\ }\href {\doibase 10.1063/5.0011516} {\bibfield
  {journal} {\bibinfo  {journal} {Physics of Plasmas}\ }\textbf {\bibinfo
  {volume} {27}},\ \bibinfo {pages} {062109} (\bibinfo {year}
  {2020}{\natexlab{a}})}\BibitemShut {NoStop}%
\bibitem [{\citenamefont {Ochs}\ and\ \citenamefont
  {Fisch}(2020{\natexlab{b}})}]{Ochs2020a}%
  \BibitemOpen
  \bibfield  {author} {\bibinfo {author} {\bibfnamefont {I.~E.}\ \bibnamefont
  {Ochs}}\ and\ \bibinfo {author} {\bibfnamefont {N.~J.}\ \bibnamefont
  {Fisch}},\ }\bibfield  {title} {\enquote {\bibinfo {title} {{Magnetogenesis
  by wave-driven momentum exchange}},}\ }\href {\doibase
  10.3847/1538-4357/abc4e8} {\bibfield  {journal} {\bibinfo  {journal} {The
  Astrophysical Journal}\ }\textbf {\bibinfo {volume} {905}} (\bibinfo {year}
  {2020}{\natexlab{b}}),\ 10.3847/1538-4357/abc4e8}\BibitemShut {NoStop}%
\bibitem [{\citenamefont {Bellan}(2008)}]{bellan2008fundamentals}%
  \BibitemOpen
  \bibfield  {author} {\bibinfo {author} {\bibfnamefont {P.~M.}\ \bibnamefont
  {Bellan}},\ }\href@noop {} {\emph {\bibinfo {title} {{Fundamentals of plasma
  physics}}}}\ (\bibinfo  {publisher} {Cambridge University Press},\ \bibinfo
  {year} {2008})\BibitemShut {NoStop}%
\bibitem [{\citenamefont {Fisch}(1978{\natexlab{b}})}]{fisch1978currentDrive}%
  \BibitemOpen
  \bibfield  {author} {\bibinfo {author} {\bibfnamefont {N.~J.}\ \bibnamefont
  {Fisch}},\ }\bibfield  {title} {\enquote {\bibinfo {title} {{Confining a
  Tokamak Plasma with rf-Driven Currents}},}\ }\href {\doibase
  10.1103/PhysRevLett.41.873} {\bibfield  {journal} {\bibinfo  {journal} {Phys.
  Rev. Lett.}\ }\textbf {\bibinfo {volume} {41}},\ \bibinfo {pages} {873--876}
  (\bibinfo {year} {1978}{\natexlab{b}})}\BibitemShut {NoStop}%
\bibitem [{\citenamefont {Fisch}\ and\ \citenamefont
  {Boozer}(1980)}]{fisch1980creating}%
  \BibitemOpen
  \bibfield  {author} {\bibinfo {author} {\bibfnamefont {N.~J.}\ \bibnamefont
  {Fisch}}\ and\ \bibinfo {author} {\bibfnamefont {A.~H.}\ \bibnamefont
  {Boozer}},\ }\bibfield  {title} {\enquote {\bibinfo {title} {{Creating an
  asymmetric plasma resistivity with waves}},}\ }\href@noop {} {\bibfield
  {journal} {\bibinfo  {journal} {Physical Review Letters}\ }\textbf {\bibinfo
  {volume} {45}},\ \bibinfo {pages} {720} (\bibinfo {year} {1980})}\BibitemShut
  {NoStop}%
\bibitem [{\citenamefont {Braginskii}(1965)}]{braginskii1965transport}%
  \BibitemOpen
  \bibfield  {author} {\bibinfo {author} {\bibfnamefont {S.~I.}\ \bibnamefont
  {Braginskii}},\ }\bibfield  {title} {\enquote {\bibinfo {title} {{Transport
  processes in a plasma}},}\ }\href@noop {} {\bibfield  {journal} {\bibinfo
  {journal} {Reviews of Plasma Physics}\ }\textbf {\bibinfo {volume} {1}},\
  \bibinfo {pages} {205} (\bibinfo {year} {1965})}\BibitemShut {NoStop}%
\bibitem [{\citenamefont {Helander}\ and\ \citenamefont
  {Sigmar}(2005)}]{helander2005collisional}%
  \BibitemOpen
  \bibfield  {author} {\bibinfo {author} {\bibfnamefont {P.}~\bibnamefont
  {Helander}}\ and\ \bibinfo {author} {\bibfnamefont {D.~J.}\ \bibnamefont
  {Sigmar}},\ }\href@noop {} {\emph {\bibinfo {title} {{Collisional transport
  in magnetized plasmas}}}},\ Vol.~\bibinfo {volume} {4}\ (\bibinfo
  {publisher} {Cambridge University Press},\ \bibinfo {year}
  {2005})\BibitemShut {NoStop}%
\bibitem [{\citenamefont {Rozhansky}(2008)}]{Rozhansky2008}%
  \BibitemOpen
  \bibfield  {author} {\bibinfo {author} {\bibfnamefont {V.}~\bibnamefont
  {Rozhansky}},\ }\bibfield  {title} {\enquote {\bibinfo {title} {{Mechanisma
  of transverse conductivity and generation of self-consistent electric fields
  in strongly ionized magnetized plasma}},}\ }in\ \href@noop {} {\emph
  {\bibinfo {booktitle} {Review of Plasma Physics}}}\ (\bibinfo  {publisher}
  {Springer},\ \bibinfo {year} {2008})\ pp.\ \bibinfo {pages}
  {1--52}\BibitemShut {NoStop}%
\bibitem [{\citenamefont {Ochs}\ and\ \citenamefont
  {Fisch}(2018)}]{ochs2018favorable}%
  \BibitemOpen
  \bibfield  {author} {\bibinfo {author} {\bibfnamefont {I.~E.}\ \bibnamefont
  {Ochs}}\ and\ \bibinfo {author} {\bibfnamefont {N.~J.}\ \bibnamefont
  {Fisch}},\ }\bibfield  {title} {\enquote {\bibinfo {title} {{Favorable
  Collisional Demixing of Ash and Fuel in Magnetized Inertial Fusion}},}\
  }\href {\doibase 10.1103/PhysRevLett.121.235002} {\bibfield  {journal}
  {\bibinfo  {journal} {Phys. Rev. Lett.}\ }\textbf {\bibinfo {volume} {121}},\
  \bibinfo {pages} {235002} (\bibinfo {year} {2018})}\BibitemShut {NoStop}%
\bibitem [{\citenamefont {Kolmes}\ \emph {et~al.}(2019)\citenamefont {Kolmes},
  \citenamefont {Ochs}, \citenamefont {Mlodik}, \citenamefont {Rax},
  \citenamefont {Gueroult},\ and\ \citenamefont {Fisch}}]{kolmes2019radial}%
  \BibitemOpen
  \bibfield  {author} {\bibinfo {author} {\bibfnamefont {E.~J.}\ \bibnamefont
  {Kolmes}}, \bibinfo {author} {\bibfnamefont {I.~E.}\ \bibnamefont {Ochs}},
  \bibinfo {author} {\bibfnamefont {M.~E.}\ \bibnamefont {Mlodik}}, \bibinfo
  {author} {\bibfnamefont {J.~M.}\ \bibnamefont {Rax}}, \bibinfo {author}
  {\bibfnamefont {R.}~\bibnamefont {Gueroult}}, \ and\ \bibinfo {author}
  {\bibfnamefont {N.~J.}\ \bibnamefont {Fisch}},\ }\bibfield  {title} {\enquote
  {\bibinfo {title} {{Radial current and rotation profile tailoring in highly
  ionized linear plasma devices}},}\ }\href {\doibase 10.1063/1.5115788}
  {\bibfield  {journal} {\bibinfo  {journal} {Physics of Plasmas}\ }\textbf
  {\bibinfo {volume} {26}} (\bibinfo {year} {2019}),\
  10.1063/1.5115788}\BibitemShut {NoStop}%
\bibitem [{\citenamefont {Chankin}\ and\ \citenamefont
  {Stangeby}(1994)}]{Chankin1994}%
  \BibitemOpen
  \bibfield  {author} {\bibinfo {author} {\bibfnamefont {A.~V.}\ \bibnamefont
  {Chankin}}\ and\ \bibinfo {author} {\bibfnamefont {P.~C.}\ \bibnamefont
  {Stangeby}},\ }\bibfield  {title} {\enquote {\bibinfo {title} {{The effect of
  diamagnetic drift on the boundary conditions in tokamak scrape-off layers and
  the distribution of plasma fluxes near the target}},}\ }\href {\doibase
  10.1088/0741-3335/36/9/008} {\bibfield  {journal} {\bibinfo  {journal}
  {Plasma Physics and Controlled Fusion}\ }\textbf {\bibinfo {volume} {36}},\
  \bibinfo {pages} {1485--1499} (\bibinfo {year} {1994})}\BibitemShut {NoStop}%
\bibitem [{\citenamefont {Chankin}\ and\ \citenamefont
  {Stangeby}(1996)}]{Chankin1996}%
  \BibitemOpen
  \bibfield  {author} {\bibinfo {author} {\bibfnamefont {A.~V.}\ \bibnamefont
  {Chankin}}\ and\ \bibinfo {author} {\bibfnamefont {P.~C.}\ \bibnamefont
  {Stangeby}},\ }\bibfield  {title} {\enquote {\bibinfo {title} {{Toroidicity
  in the tokamak SOL: effects on poloidal asymmetries, radial current and the L
  - H transition}},}\ }\href {\doibase 10.1088/0741-3335/38/11/002} {\bibfield
  {journal} {\bibinfo  {journal} {Plasma Physics and Controlled Fusion}\
  }\textbf {\bibinfo {volume} {38}},\ \bibinfo {pages} {1879--1903} (\bibinfo
  {year} {1996})}\BibitemShut {NoStop}%
\bibitem [{\citenamefont {Stoltzfus-Dueck}(2019)}]{Stoltzfus-Dueck2019}%
  \BibitemOpen
  \bibfield  {author} {\bibinfo {author} {\bibfnamefont {T.}~\bibnamefont
  {Stoltzfus-Dueck}},\ }\bibfield  {title} {\enquote {\bibinfo {title}
  {{Intrinsic rotation in axisymmetric devices}},}\ }\href {\doibase
  10.1088/1361-6587/ab4376} {\bibfield  {journal} {\bibinfo  {journal} {Plasma
  Physics and Controlled Fusion}\ }\textbf {\bibinfo {volume} {61}} (\bibinfo
  {year} {2019}),\ 10.1088/1361-6587/ab4376}\BibitemShut {NoStop}%
\bibitem [{\citenamefont {Rozhansky}\ and\ \citenamefont
  {Tendler}(1994)}]{Rozhansky1994}%
  \BibitemOpen
  \bibfield  {author} {\bibinfo {author} {\bibfnamefont {V.}~\bibnamefont
  {Rozhansky}}\ and\ \bibinfo {author} {\bibfnamefont {M.}~\bibnamefont
  {Tendler}},\ }\bibfield  {title} {\enquote {\bibinfo {title} {{The impact of
  a biasing radial electric field on the scrape‐off layer in a divertor
  tokamak}},}\ }\href {\doibase 10.1063/1.870598} {\bibfield  {journal}
  {\bibinfo  {journal} {Physics of Plasmas}\ }\textbf {\bibinfo {volume} {1}},\
  \bibinfo {pages} {2711--2717} (\bibinfo {year} {1994})}\BibitemShut {NoStop}%
\bibitem [{\citenamefont {Ochs}\ and\ \citenamefont
  {Fisch}()}]{ochs2021nonresonant}%
  \BibitemOpen
  \bibfield  {author} {\bibinfo {author} {\bibfnamefont {I.~E.}\ \bibnamefont
  {Ochs}}\ and\ \bibinfo {author} {\bibfnamefont {N.~J.}\ \bibnamefont
  {Fisch}},\ }\bibfield  {title} {\enquote {\bibinfo {title} {{Nonresonant
  Diffusion in Alpha Channeling}},}\ }\href@noop {} {\bibfield  {journal}
  {\bibinfo  {journal} {Physical Review Letters}\ }\textbf {\bibinfo {volume}
  {In Press}},\ \bibinfo {pages} {Arxiv: physics.plasm--ph/2012.06532}},\
  \Eprint {http://arxiv.org/abs/2012.06532} {arXiv:2012.06532
  [physics.plasm-ph]} \BibitemShut {NoStop}%
\bibitem [{\citenamefont {Berk}\ and\ \citenamefont {Molvig}(1983)}]{Berk1983}%
  \BibitemOpen
  \bibfield  {author} {\bibinfo {author} {\bibfnamefont {H.~L.}\ \bibnamefont
  {Berk}}\ and\ \bibinfo {author} {\bibfnamefont {K.}~\bibnamefont {Molvig}},\
  }\bibfield  {title} {\enquote {\bibinfo {title} {{Nonintrinsic ambipolar
  diffusion in turbulence theory}},}\ }\href {\doibase 10.1063/1.864325}
  {\bibfield  {journal} {\bibinfo  {journal} {Physics of Fluids}\ }\textbf
  {\bibinfo {volume} {26}},\ \bibinfo {pages} {1385--1388} (\bibinfo {year}
  {1983})}\BibitemShut {NoStop}%
\bibitem [{\citenamefont {Diamond}\ and\ \citenamefont
  {Kim}(1991)}]{Diamond1991}%
  \BibitemOpen
  \bibfield  {author} {\bibinfo {author} {\bibfnamefont {P.~H.}\ \bibnamefont
  {Diamond}}\ and\ \bibinfo {author} {\bibfnamefont {Y.~B.}\ \bibnamefont
  {Kim}},\ }\bibfield  {title} {\enquote {\bibinfo {title} {{Theory of mean
  poloidal flow generation by turbulence}},}\ }\href {\doibase
  10.1063/1.859681} {\bibfield  {journal} {\bibinfo  {journal} {Physics of
  Fluids B}\ }\textbf {\bibinfo {volume} {3}},\ \bibinfo {pages} {1626--1633}
  (\bibinfo {year} {1991})}\BibitemShut {NoStop}%
\bibitem [{\citenamefont {Diamond}\ \emph {et~al.}(2008)\citenamefont
  {Diamond}, \citenamefont {McDevitt}, \citenamefont {G{\"{u}}rcan},
  \citenamefont {Hahm},\ and\ \citenamefont {Naulin}}]{Diamond2008}%
  \BibitemOpen
  \bibfield  {author} {\bibinfo {author} {\bibfnamefont {P.~H.}\ \bibnamefont
  {Diamond}}, \bibinfo {author} {\bibfnamefont {C.~J.}\ \bibnamefont
  {McDevitt}}, \bibinfo {author} {\bibfnamefont {{\"{O}}.~D.}\ \bibnamefont
  {G{\"{u}}rcan}}, \bibinfo {author} {\bibfnamefont {T.~S.}\ \bibnamefont
  {Hahm}}, \ and\ \bibinfo {author} {\bibfnamefont {V.}~\bibnamefont
  {Naulin}},\ }\bibfield  {title} {\enquote {\bibinfo {title} {{Transport of
  parallel momentum by collisionless drift wave turbulence}},}\ }\href
  {\doibase 10.1063/1.2826436} {\bibfield  {journal} {\bibinfo  {journal}
  {Physics of Plasmas}\ }\textbf {\bibinfo {volume} {15}} (\bibinfo {year}
  {2008}),\ 10.1063/1.2826436}\BibitemShut {NoStop}%
\bibitem [{\citenamefont {Krommes}\ and\ \citenamefont
  {Hammet}(2013)}]{Krommes2013}%
  \BibitemOpen
  \bibfield  {author} {\bibinfo {author} {\bibfnamefont {J.~A.}\ \bibnamefont
  {Krommes}}\ and\ \bibinfo {author} {\bibfnamefont {G.~W.}\ \bibnamefont
  {Hammet}},\ }\href@noop {} {\enquote {\bibinfo {title} {{Report of the Study
  Group GK2 on Momentum Transport in Gyrokinetics}},}\ }\bibinfo {type} {Tech.
  Rep.}\ (\bibinfo  {institution} {PPPL Report PPPL-4945},\ \bibinfo {year}
  {2013})\BibitemShut {NoStop}%
\bibitem [{\citenamefont {Berry}, \citenamefont {Jaeger},\ and\ \citenamefont
  {Batchelor}(1999)}]{Berry1999}%
  \BibitemOpen
  \bibfield  {author} {\bibinfo {author} {\bibfnamefont {L.~A.}\ \bibnamefont
  {Berry}}, \bibinfo {author} {\bibfnamefont {E.~F.}\ \bibnamefont {Jaeger}}, \
  and\ \bibinfo {author} {\bibfnamefont {D.~B.}\ \bibnamefont {Batchelor}},\
  }\bibfield  {title} {\enquote {\bibinfo {title} {{Wave-induced momentum
  transport and flow drive in tokamak plasmas}},}\ }\href {\doibase
  10.1103/PhysRevLett.82.1871} {\bibfield  {journal} {\bibinfo  {journal}
  {Physical Review Letters}\ }\textbf {\bibinfo {volume} {82}},\ \bibinfo
  {pages} {1871--1874} (\bibinfo {year} {1999})}\BibitemShut {NoStop}%
\bibitem [{\citenamefont {Jaeger}, \citenamefont {Berry},\ and\ \citenamefont
  {Batchelor}(2000)}]{Jaeger2000}%
  \BibitemOpen
  \bibfield  {author} {\bibinfo {author} {\bibfnamefont {E.~F.}\ \bibnamefont
  {Jaeger}}, \bibinfo {author} {\bibfnamefont {L.~A.}\ \bibnamefont {Berry}}, \
  and\ \bibinfo {author} {\bibfnamefont {D.~B.}\ \bibnamefont {Batchelor}},\
  }\bibfield  {title} {\enquote {\bibinfo {title} {{Second-order radio
  frequency kinetic theory with applications to flow drive and heating in
  tokamak plasmas}},}\ }\href {\doibase 10.1063/1.873868} {\bibfield  {journal}
  {\bibinfo  {journal} {Physics of Plasmas}\ }\textbf {\bibinfo {volume} {7}},\
  \bibinfo {pages} {641--656} (\bibinfo {year} {2000})}\BibitemShut {NoStop}%
\bibitem [{\citenamefont {Myra}\ and\ \citenamefont
  {D'Ippolito}(2000)}]{Myra2000}%
  \BibitemOpen
  \bibfield  {author} {\bibinfo {author} {\bibfnamefont {J.~R.}\ \bibnamefont
  {Myra}}\ and\ \bibinfo {author} {\bibfnamefont {D.~A.}\ \bibnamefont
  {D'Ippolito}},\ }\bibfield  {title} {\enquote {\bibinfo {title} {{Poloidal
  force generation by applied radio frequency waves}},}\ }\href {\doibase
  10.1063/1.1286865} {\bibfield  {journal} {\bibinfo  {journal} {Physics of
  Plasmas}\ }\textbf {\bibinfo {volume} {7}},\ \bibinfo {pages} {3600--3609}
  (\bibinfo {year} {2000})}\BibitemShut {NoStop}%
\bibitem [{\citenamefont {Myra}\ and\ \citenamefont
  {D'Ippolito}(2002)}]{Myra2002}%
  \BibitemOpen
  \bibfield  {author} {\bibinfo {author} {\bibfnamefont {J.~R.}\ \bibnamefont
  {Myra}}\ and\ \bibinfo {author} {\bibfnamefont {D.~A.}\ \bibnamefont
  {D'Ippolito}},\ }\bibfield  {title} {\enquote {\bibinfo {title} {{Toroidal
  formulation of nonlinear-rf-driven flows}},}\ }\href {\doibase
  10.1063/1.1496762} {\bibfield  {journal} {\bibinfo  {journal} {Physics of
  Plasmas}\ }\textbf {\bibinfo {volume} {9}},\ \bibinfo {pages} {3867}
  (\bibinfo {year} {2002})}\BibitemShut {NoStop}%
\bibitem [{\citenamefont {Myra}\ \emph {et~al.}(2004)\citenamefont {Myra},
  \citenamefont {Berry}, \citenamefont {D'Lppolito},\ and\ \citenamefont
  {Jaeger}}]{Myra2004}%
  \BibitemOpen
  \bibfield  {author} {\bibinfo {author} {\bibfnamefont {J.~R.}\ \bibnamefont
  {Myra}}, \bibinfo {author} {\bibfnamefont {L.~A.}\ \bibnamefont {Berry}},
  \bibinfo {author} {\bibfnamefont {D.~A.}\ \bibnamefont {D'Lppolito}}, \ and\
  \bibinfo {author} {\bibfnamefont {E.~F.}\ \bibnamefont {Jaeger}},\ }\bibfield
   {title} {\enquote {\bibinfo {title} {{Nonlinear fluxes and forces from
  radio-frequency waves with application to driven flows in tokamaks}},}\
  }\href {\doibase 10.1063/1.1690298} {\bibfield  {journal} {\bibinfo
  {journal} {Physics of Plasmas}\ }\textbf {\bibinfo {volume} {11}},\ \bibinfo
  {pages} {1786--1798} (\bibinfo {year} {2004})}\BibitemShut {NoStop}%
\bibitem [{\citenamefont {Chen}\ and\ \citenamefont {Gao}(2013)}]{Chen2013}%
  \BibitemOpen
  \bibfield  {author} {\bibinfo {author} {\bibfnamefont {J.}~\bibnamefont
  {Chen}}\ and\ \bibinfo {author} {\bibfnamefont {Z.}~\bibnamefont {Gao}},\
  }\bibfield  {title} {\enquote {\bibinfo {title} {{Second-order radio
  frequency kinetic theory revisited: Resolving inconsistency with conventional
  fluid theory}},}\ }\href {\doibase 10.1063/1.4817812} {\bibfield  {journal}
  {\bibinfo  {journal} {Physics of Plasmas}\ }\textbf {\bibinfo {volume}
  {20}},\ \bibinfo {pages} {082508} (\bibinfo {year} {2013})}\BibitemShut
  {NoStop}%
\bibitem [{\citenamefont {Lee}\ \emph {et~al.}(2012)\citenamefont {Lee},
  \citenamefont {Parra}, \citenamefont {Parker},\ and\ \citenamefont
  {Bonoli}}]{Lee2012}%
  \BibitemOpen
  \bibfield  {author} {\bibinfo {author} {\bibfnamefont {J.}~\bibnamefont
  {Lee}}, \bibinfo {author} {\bibfnamefont {F.~I.}\ \bibnamefont {Parra}},
  \bibinfo {author} {\bibfnamefont {R.~R.}\ \bibnamefont {Parker}}, \ and\
  \bibinfo {author} {\bibfnamefont {P.~T.}\ \bibnamefont {Bonoli}},\ }\bibfield
   {title} {\enquote {\bibinfo {title} {{Perpendicular momentum injection by
  lower hybrid wave in a tokamak}},}\ }\href {\doibase
  10.1088/0741-3335/54/12/125005} {\bibfield  {journal} {\bibinfo  {journal}
  {Plasma Physics and Controlled Fusion}\ }\textbf {\bibinfo {volume} {54}}
  (\bibinfo {year} {2012}),\ 10.1088/0741-3335/54/12/125005}\BibitemShut
  {NoStop}%
\bibitem [{\citenamefont {Guan}\ \emph
  {et~al.}(2013{\natexlab{a}})\citenamefont {Guan}, \citenamefont {Dodin},
  \citenamefont {Qin}, \citenamefont {Liu},\ and\ \citenamefont
  {Fisch}}]{guan2013plasma}%
  \BibitemOpen
  \bibfield  {author} {\bibinfo {author} {\bibfnamefont {X.}~\bibnamefont
  {Guan}}, \bibinfo {author} {\bibfnamefont {I.~Y.}\ \bibnamefont {Dodin}},
  \bibinfo {author} {\bibfnamefont {H.}~\bibnamefont {Qin}}, \bibinfo {author}
  {\bibfnamefont {J.}~\bibnamefont {Liu}}, \ and\ \bibinfo {author}
  {\bibfnamefont {N.~J.}\ \bibnamefont {Fisch}},\ }\bibfield  {title} {\enquote
  {\bibinfo {title} {{On plasma rotation induced by waves in tokamaks}},}\
  }\href@noop {} {\bibfield  {journal} {\bibinfo  {journal} {Physics of
  Plasmas}\ }\textbf {\bibinfo {volume} {20}},\ \bibinfo {pages} {102105}
  (\bibinfo {year} {2013}{\natexlab{a}})}\BibitemShut {NoStop}%
\bibitem [{\citenamefont {Guan}\ \emph
  {et~al.}(2013{\natexlab{b}})\citenamefont {Guan}, \citenamefont {Qin},
  \citenamefont {Liu},\ and\ \citenamefont {Fisch}}]{guan2013toroidal}%
  \BibitemOpen
  \bibfield  {author} {\bibinfo {author} {\bibfnamefont {X.}~\bibnamefont
  {Guan}}, \bibinfo {author} {\bibfnamefont {H.}~\bibnamefont {Qin}}, \bibinfo
  {author} {\bibfnamefont {J.}~\bibnamefont {Liu}}, \ and\ \bibinfo {author}
  {\bibfnamefont {N.~J.}\ \bibnamefont {Fisch}},\ }\bibfield  {title} {\enquote
  {\bibinfo {title} {{On the toroidal plasma rotations induced by lower hybrid
  waves}},}\ }\href@noop {} {\bibfield  {journal} {\bibinfo  {journal} {Physics
  of Plasmas}\ }\textbf {\bibinfo {volume} {20}},\ \bibinfo {pages} {022502}
  (\bibinfo {year} {2013}{\natexlab{b}})}\BibitemShut {NoStop}%
\bibitem [{\citenamefont {Kennel}\ and\ \citenamefont
  {Engelmann}(1966)}]{kennel1966magnetizedQL}%
  \BibitemOpen
  \bibfield  {author} {\bibinfo {author} {\bibfnamefont {C.~F.}\ \bibnamefont
  {Kennel}}\ and\ \bibinfo {author} {\bibfnamefont {F.}~\bibnamefont
  {Engelmann}},\ }\bibfield  {title} {\enquote {\bibinfo {title} {{Velocity
  Space Diffusion from Weak Plasma Turbulence in a Magnetic Field}},}\ }\href
  {\doibase 10.1063/1.1761629} {\bibfield  {journal} {\bibinfo  {journal} {The
  Physics of Fluids}\ }\textbf {\bibinfo {volume} {9}},\ \bibinfo {pages}
  {2377--2388} (\bibinfo {year} {1966})}\BibitemShut {NoStop}%
\bibitem [{\citenamefont {Dodin}\ and\ \citenamefont
  {Fisch}(2012)}]{dodin2012axiomatic}%
  \BibitemOpen
  \bibfield  {author} {\bibinfo {author} {\bibfnamefont {I.~Y.}\ \bibnamefont
  {Dodin}}\ and\ \bibinfo {author} {\bibfnamefont {N.~J.}\ \bibnamefont
  {Fisch}},\ }\bibfield  {title} {\enquote {\bibinfo {title} {{Axiomatic
  geometrical optics, Abraham-Minkowski controversy, and photon properties
  derived classically}},}\ }\href@noop {} {\bibfield  {journal} {\bibinfo
  {journal} {Physical Review A}\ }\textbf {\bibinfo {volume} {86}},\ \bibinfo
  {pages} {53834} (\bibinfo {year} {2012})}\BibitemShut {NoStop}%
\bibitem [{\citenamefont {Brambilla}(1976)}]{Brambilla1976}%
  \BibitemOpen
  \bibfield  {author} {\bibinfo {author} {\bibfnamefont {M.}~\bibnamefont
  {Brambilla}},\ }\bibfield  {title} {\enquote {\bibinfo {title} {{Slow-wave
  launching at the lower hybrid frequency using a phased waveguide array}},}\
  }\href {\doibase 10.1088/0029-5515/16/1/005} {\bibfield  {journal} {\bibinfo
  {journal} {Nuclear Fusion}\ }\textbf {\bibinfo {volume} {16}},\ \bibinfo
  {pages} {47--54} (\bibinfo {year} {1976})}\BibitemShut {NoStop}%
\bibitem [{\citenamefont {Brambilla}(1979)}]{Brambilla1979}%
  \BibitemOpen
  \bibfield  {author} {\bibinfo {author} {\bibfnamefont {M.}~\bibnamefont
  {Brambilla}},\ }\bibfield  {title} {\enquote {\bibinfo {title} {{Waveguide
  launching of lower hybrid waves}},}\ }\href {\doibase
  10.1088/0029-5515/19/10/006} {\bibfield  {journal} {\bibinfo  {journal}
  {Nuclear Fusion}\ }\textbf {\bibinfo {volume} {19}},\ \bibinfo {pages}
  {1343--1357} (\bibinfo {year} {1979})}\BibitemShut {NoStop}%
\bibitem [{\citenamefont {Carroll}(2003)}]{carroll2003spacetime}%
  \BibitemOpen
  \bibfield  {author} {\bibinfo {author} {\bibfnamefont {S.~M.}\ \bibnamefont
  {Carroll}},\ }\href@noop {} {\emph {\bibinfo {title} {{Spacetime and
  geometry. An introduction to general relativity}}}},\ Vol.~\bibinfo {volume}
  {1}\ (\bibinfo  {publisher} {Pearson},\ \bibinfo {year} {2003})\BibitemShut
  {NoStop}%
\bibitem [{\citenamefont {Blandford}\ and\ \citenamefont
  {Thorne}(2017)}]{Blandford2017}%
  \BibitemOpen
  \bibfield  {author} {\bibinfo {author} {\bibfnamefont {R.~D.}\ \bibnamefont
  {Blandford}}\ and\ \bibinfo {author} {\bibfnamefont {K.~S.}\ \bibnamefont
  {Thorne}},\ }\href@noop {} {\emph {\bibinfo {title} {{Modern Classical
  Physics: Optics, Fluids, Plasmas, Elasticity, Relativity, and Statistical
  Physics}}}}\ (\bibinfo  {publisher} {Princeton University Press},\ \bibinfo
  {address} {Princeton, NJ},\ \bibinfo {year} {2017})\BibitemShut {NoStop}%
\bibitem [{\citenamefont {Tracy}\ \emph {et~al.}(2014)\citenamefont {Tracy},
  \citenamefont {Brizard}, \citenamefont {Richardson},\ and\ \citenamefont
  {Kaufman}}]{Tracy2014}%
  \BibitemOpen
  \bibfield  {author} {\bibinfo {author} {\bibfnamefont {E.~R.}\ \bibnamefont
  {Tracy}}, \bibinfo {author} {\bibfnamefont {A.~J.}\ \bibnamefont {Brizard}},
  \bibinfo {author} {\bibfnamefont {A.~S.}\ \bibnamefont {Richardson}}, \ and\
  \bibinfo {author} {\bibfnamefont {A.~N.}\ \bibnamefont {Kaufman}},\
  }\href@noop {} {\emph {\bibinfo {title} {{Ray tracing and beyond: phase space
  methods in plasma wave theory}}}}\ (\bibinfo  {publisher} {Cambridge
  University Press},\ \bibinfo {year} {2014})\BibitemShut {NoStop}%
\bibitem [{\citenamefont {Kaufman}(1972)}]{Kaufman1972a}%
  \BibitemOpen
  \bibfield  {author} {\bibinfo {author} {\bibfnamefont {A.~N.}\ \bibnamefont
  {Kaufman}},\ }\bibfield  {title} {\enquote {\bibinfo {title} {{Reformulation
  of quasi-linear theory}},}\ }\href@noop {} {\bibfield  {journal} {\bibinfo
  {journal} {J. Plasma Physics}\ }\textbf {\bibinfo {volume} {8}},\ \bibinfo
  {pages} {1--5} (\bibinfo {year} {1972})}\BibitemShut {NoStop}%
\bibitem [{\citenamefont {Stix}(1965)}]{Stix1965}%
  \BibitemOpen
  \bibfield  {author} {\bibinfo {author} {\bibfnamefont {T.~H.}\ \bibnamefont
  {Stix}},\ }\bibfield  {title} {\enquote {\bibinfo {title} {{Radiation and
  absorption via mode conversion in an inhomogeneous collision-free plasma}},}\
  }\href {\doibase 10.1103/PhysRevLett.15.878} {\bibfield  {journal} {\bibinfo
  {journal} {Physical Review Letters}\ }\textbf {\bibinfo {volume} {15}},\
  \bibinfo {pages} {878--882} (\bibinfo {year} {1965})}\BibitemShut {NoStop}%
\bibitem [{\citenamefont {Griffiths}(2017)}]{Griffiths2017}%
  \BibitemOpen
  \bibfield  {author} {\bibinfo {author} {\bibfnamefont {D.~J.}\ \bibnamefont
  {Griffiths}},\ }\href@noop {} {\emph {\bibinfo {title} {{Introduction to
  Electrodynamics}}}},\ \bibinfo {edition} {4th}\ ed.\ (\bibinfo  {publisher}
  {Cambridge University Press},\ \bibinfo {address} {Cambridge},\ \bibinfo
  {year} {2017})\BibitemShut {NoStop}%
\bibitem [{\citenamefont {Jackson}(1999)}]{Jackson1999}%
  \BibitemOpen
  \bibfield  {author} {\bibinfo {author} {\bibfnamefont {J.~D.}\ \bibnamefont
  {Jackson}},\ }\href@noop {} {\emph {\bibinfo {title} {{Classical
  electrodynamics}}}},\ \bibinfo {edition} {3rd}\ ed.\ (\bibinfo  {publisher}
  {John Wiley and Sons},\ \bibinfo {year} {1999})\BibitemShut {NoStop}%
\bibitem [{\citenamefont {Gao}, \citenamefont {Fisch},\ and\ \citenamefont
  {Qin}(2006)}]{Gao2006}%
  \BibitemOpen
  \bibfield  {author} {\bibinfo {author} {\bibfnamefont {Z.}~\bibnamefont
  {Gao}}, \bibinfo {author} {\bibfnamefont {N.~J.}\ \bibnamefont {Fisch}}, \
  and\ \bibinfo {author} {\bibfnamefont {H.}~\bibnamefont {Qin}},\ }\bibfield
  {title} {\enquote {\bibinfo {title} {{Nonlinear ponderomotive force by low
  frequency waves and nonresonant current drive}},}\ }\href {\doibase
  10.1063/1.2397584} {\bibfield  {journal} {\bibinfo  {journal} {Physics of
  Plasmas}\ }\textbf {\bibinfo {volume} {13}},\ \bibinfo {pages} {112307}
  (\bibinfo {year} {2006})}\BibitemShut {NoStop}%
\bibitem [{\citenamefont {Gao}\ \emph {et~al.}(2007)\citenamefont {Gao},
  \citenamefont {Fisch}, \citenamefont {Qin},\ and\ \citenamefont
  {Myra}}]{Gao2007}%
  \BibitemOpen
  \bibfield  {author} {\bibinfo {author} {\bibfnamefont {Z.}~\bibnamefont
  {Gao}}, \bibinfo {author} {\bibfnamefont {N.~J.}\ \bibnamefont {Fisch}},
  \bibinfo {author} {\bibfnamefont {H.}~\bibnamefont {Qin}}, \ and\ \bibinfo
  {author} {\bibfnamefont {J.~R.}\ \bibnamefont {Myra}},\ }\bibfield  {title}
  {\enquote {\bibinfo {title} {{Nonlinear nonresonant forces by radio-frequency
  waves in plasmas}},}\ }\href {\doibase 10.1063/1.2775431} {\bibfield
  {journal} {\bibinfo  {journal} {Physics of Plasmas}\ }\textbf {\bibinfo
  {volume} {14}} (\bibinfo {year} {2007}),\ 10.1063/1.2775431}\BibitemShut
  {NoStop}%
\bibitem [{\citenamefont {Verdon}\ \emph {et~al.}(2009)\citenamefont {Verdon},
  \citenamefont {Cairns}, \citenamefont {Melrose},\ and\ \citenamefont
  {Robinson}}]{Verdon2009}%
  \BibitemOpen
  \bibfield  {author} {\bibinfo {author} {\bibfnamefont {A.~L.}\ \bibnamefont
  {Verdon}}, \bibinfo {author} {\bibfnamefont {I.~H.}\ \bibnamefont {Cairns}},
  \bibinfo {author} {\bibfnamefont {D.~B.}\ \bibnamefont {Melrose}}, \ and\
  \bibinfo {author} {\bibfnamefont {P.~A.}\ \bibnamefont {Robinson}},\
  }\bibfield  {title} {\enquote {\bibinfo {title} {{Warm electromagnetic lower
  hybrid wave dispersion relation}},}\ }\href {\doibase 10.1063/1.3132628}
  {\bibfield  {journal} {\bibinfo  {journal} {Physics of Plasmas}\ }\textbf
  {\bibinfo {volume} {16}},\ \bibinfo {pages} {052105} (\bibinfo {year}
  {2009})}\BibitemShut {NoStop}%
\bibitem [{\citenamefont {Dewar}(1973)}]{Dewar1973}%
  \BibitemOpen
  \bibfield  {author} {\bibinfo {author} {\bibfnamefont {R.~L.}\ \bibnamefont
  {Dewar}},\ }\bibfield  {title} {\enquote {\bibinfo {title} {{Oscillation
  center quasilinear theory}},}\ }\href {\doibase 10.1063/1.1694473} {\bibfield
   {journal} {\bibinfo  {journal} {Physics of Fluids}\ }\textbf {\bibinfo
  {volume} {16}},\ \bibinfo {pages} {1102--1107} (\bibinfo {year}
  {1973})}\BibitemShut {NoStop}%
\end{thebibliography}
